\documentclass[a4paper,11pt]{article}
\pdfoutput=1

\usepackage{jinstpub}
\usepackage{siunitx}
\title{An improved trigger for Askaryan radio detectors}

\author[a,b]{Christian Glaser}
\author[b]{Steven~W. Barwick}

\affiliation[a]{Uppsala University Department of Physics and Astronomy, Uppsala, SE-752
37, Sweden}
\affiliation[b]{Department of Physics and Astronomy, University of California, Irvine, CA 92697, USA}
\emailAdd{christian.glaser@physics.uu.se}
\emailAdd{sbarwick@uci.edu}

\abstract{
High-energy neutrinos with energies above a few \SI{e16}{eV} can be measured efficiently with in-ice radio detectors which complement optical detectors such as IceCube at higher energies. Several pilot arrays explore the radio technology successfully in Antarctica. Because of the low flux and interaction cross-section of neutrinos it is vital to increase the sensitivity of the radio detector as much as possible. In this manuscript, different approaches to trigger on high-energy neutrinos are systematically studied and optimized. We find that the sensitivity can be improved substantially (by more than 50\% between \SI{e17}{eV} and \SI{e18}{eV}) by simply restricting the bandwidth in the trigger to frequencies between \SI{80}{MHz} and \SI{200}{MHz} instead of the currently used \SI{80}{MHz} to $\sim\SI{1}{GHz}$ bandwidth. We also compare different trigger schemes that are currently being used (a simple amplitude threshold, a high/low threshold trigger and a power-integration trigger) and find that the scheme that performs best depends on the dispersion of the detector. These findings inform the detector design of future Askaryan detectors and can be used to increase the sensitivity to high-energy neutrinos significantly without any additional costs. The findings also apply to the phased array trigger concept.
}

\begin{document}
\maketitle

\section{Introduction}
\label{sec:intro}

Radio-detection of neutrinos is the most cost-effective and promising technique to study neutrinos at ultra-high energies (UHE), beyond those accessible to IceCube \cite{PhysRevLett.117.241101}. 
The detection of UHE neutrinos is a key to discovering and characterizing the most energetic sources in the universe \cite{DecadalWhitePaper} as well as fundamental measurements of the neutrino cross sections and flavor ratios at high energies. Their detection also probes the fundamental symmetries of Nature, and beyond-standard-model physics like sterile or unstable neutrinos and dark matter decaying into neutrinos \cite{DecadalWhitePaper2}.

The low flux of UHE neutrinos together with the low interaction cross-section makes their detection challenging and requires the instrumentation of huge volumes. An increase of the instrumented volume by two orders of magnitude for optical detectors, such as IceCube, is considered cost-prohibitive due to the attenuation and scattering of optical light in ice \cite{Aartsen:2013rt}. A cost-efficient way to instrument large volumes is via a sparse array of radio antenna stations installed, for instance, in the Antarctic or Arctic ice. A neutrino interaction in the ice generates a few-nanoseconds long radio flash via the Askaryan effect \cite{Askaryan}. The attenuation length of radio signals in cold ice is of $\mathcal{O}(\SI{1}{km}$) which allows a sparse instrumentation with independent radio detector stations. To further optimize the sensitivity to neutrinos, the detection threshold of each station is lowered as much as possible to increase the observed volume. Thus, it is vital to optimize the trigger as much as possible which is studied in this article. 

The feasibility of the radio technique is demonstrated by the two pilot arrays ARIANNA (see e.g. \cite{ARIANNA2015, ARIA}) installed on the Ross Ice Shelf and at the South Pole, and ARA at the South Pole \cite{ARA}. The hardware has proven to work reliably in the harsh Antarctic conditions, and it has been shown that the thermal and anthropogenic radio background can be rejected with high efficiency in an offline analysis \cite{Anker_2020, ARA2019}.

The ARIANNA detector consists of high-gain broadband LPDA antennas that are installed slightly below the surface. The ARA detector comprises antennas in holes at a depth of down to \SI{200}{m}. The limited diameter of \SI{15}{cm} restricts the choice of antennas to bicone antennas which have less gain and are less broadband compared to LPDAs but also introduce less dispersion for recorded signals. To further increase the sensitivity of a deep detector, several bicone antennas with small separation on a vertical string can be combined in an interferometric phased array \cite{ARAprogress} which produces a synthesised waveform with increased signal-to-noise ratio by coherently adding the signals of the antennas. 

Two different trigger schemes are currently being used. The ARIANNA trigger requires the time-domain signal to pass a positive and negative threshold within a time window of $\sim\SI{5}{ns}$. This is motivated by the bipolar nature of the Askaryan signals. Thus, requiring a high/low threshold has only little negative impact on the trigger efficiency of Askaryan signals but will significantly reduce the chance to trigger on thermal noise fluctuations \cite{Barwick2014}. The ARA detector uses a power integration trigger which is implemented through an integrating tunnel diode \cite{ARA2019}. The interferometric phased array also implements a power-integration trigger but implemented digitally on an FPGA as the waveforms are digitized in real time \cite{ARAprogress}. So far, these different trigger schemes have not been systematically compared on which trigger options leads to better performance. These questions will be addressed in this paper. 

Larger in-ice radio arrays are currently in the planning and/or construction phase. The RNO-G project will install 35 autonomous radio detector stations in central Greenland where each station will consist of a deep component with a phased array trigger system to maximize the sensitivity per station \cite{RNOGwhitepaper}. Each station will also have a surface component that consists of LPDA antennas which is advantageous for an improved event reconstruction \cite{GlaserICRC2019, ARIANNA2020Polarization, Gen2WhitePaper} and to measure cosmic rays \cite{Barwick2017} to reject potential background from high-energy muons that are produced in air showers \cite{GarciaFernandez2020}.

The ARIANNA-200 project plans to install 200 autonomous radio detector stations on the Ross Ice Shelf, Antarctica which would uniquely survey the vast majority of the southern sky at any instant in time due to the reflective properties of the ice-water interface at the bottom of the ice shelf \cite{ARIANNA200}. 
A radio component is also part of suggested IceCube-Gen2 detector to increase the energy reach towards higher energies \cite{Gen2WhitePaper}. 

The large number of stations and the large spacing of $>\SI{1}{km}$ between stations of future experiments require an autonomous power source, i.e., solar and wind power \cite{WindTurbineICRC2019} which limits the available power especially during the polar night.  This puts strong demands on the maximum power consumption, and real-time digitization up to GHz frequencies of all channels is considered unfeasible with currently available technology which limits available trigger algorithms. Therefore, either a (low-power) analog trigger is used (e.g. the high/low trigger of ARIANNA) or the real-time digitization is limited to the channels of the phased array with a reduced sampling rate as done for RNO-G \cite{RNOGwhitepaper}. Furthermore, to optimize the overall sensitivity, the separation between stations is increased to minimize the number of events seen in two or more stations. Therefore, each radio detector station is required to trigger autonomously. 
Thus, the limiting factor for the achievable sensitivity is the first trigger stage -- which is studied in this article -- after which the signal of all channels is being digitized. The development of second-level trigger stages that reduce the level one trigger rate by using the radio station's computing resources are currently under development but beyond the scope of this paper.

\section{Method}
\label{sec:method}
In this section, a method is presented to compare different trigger schemes and different detector responses. We will use this method in the next two sections to compare the performance of different trigger schemes and to compare the impact of trigger bandwidth on the neutrino sensitivity.

We exploit the property of Askaryan detectors that the trigger threshold is determined by the maximum data rate the detector can handle. Because the flux of high-energy neutrinos is low, the detection thresholds are decreased as much as possible to maximize the sensitivity for neutrino detection. The environment in the remote places of an Askaryan detector (e.g. the South Pole,  the Ross Ice Shelf or Greenland) is so radio quiet that thermal noise fluctuations dominate the trigger rate for most of the time (see. e.g. \cite{ARIANNA2015}). Thus, for a given trigger rate, the threshold can be determined by simulating the trigger rate on thermal noise. We note that it is of course important to keep the system noise temperature as low as possible to increase the sensitivity of a radio detector, as the RMS noise scales with the square root of the noise temperature. Throughout this analysis, we assume the same noise temperature of \SI{250}{K} for all considered setups.

The data rate is limited by the digitization/readout speed of the hardware, local storage and/or communication bandwidth and can be as high as \SI{100}{Hz} which we will assume in the following. We note that we will only consider the first trigger stage in this paper. It is advisable to further reduce the trigger rate by one or several additional trigger stages and different strategies are currently being studied. One promising approach is to use a neural network to reject thermal noise fluctuations to reduce the data rate by several orders of magnitude. This will reduce the burden on communication bandwidth, storage and computing resources significantly. 

One relevant quantity of the detector response is the dispersion introduced by the antenna response, the filters and amplifiers, or in general any component of the signal chain. 
To keep the discussion as general as possible -- to be able to derive a general dependence -- a linear dispersion of \SI{0}{ns/GHz}, \SI{10}{ns/GHz}, \SI{20}{ns/GHz} and \SI{50}{ns/GHz} is studied first. Later, the analysis is also applied to the antenna responses of an LDPA and a bicone antenna which will confirm that the general conclusions of the toy model will also be valid for a realistic detector response. 

We study three different trigger schemes
\begin{itemize}
    \item simple threshold trigger: an event is recorded if the amplitude is above or below a threshold $T$. In mathematical terms, the trigger is fulfilled if $|A_i| > T$ for any time bin $i$ where $A_i$ is the amplitude at the time bin $i$. 
    \item high/low trigger: the trigger is fulfilled if the amplitude is both above and below a threshold $T$ within a time window $\Delta t$. This trigger scheme exploits the property that the electric field of an Askaryan pulse is nearly bipolar, and the maximum positive and negative amplitudes become more symmetric after antenna and amplifier response functions are considered, whereas thermal noise fluctuations are often only one-sided. This trigger scheme is used in the ARIANNA detector \cite{ARIANNA2015}.
    \item power integration trigger: the trigger is fulfilled if the time-integrated power exceeds a certain threshold $T$. In mathematical terms: The trigger is fulfilled if $\sum\limits_{i=t_0}^{t_0 + \Delta t} A_i^2 > T$ where $\Delta t$ is the integration window, and $t_0$ is the start time of the integration. This trigger scheme is used by the ARA detector \cite{ARA} and the interferometric phased array \cite{ARAprogress}.
\end{itemize}

The second relevant quantity of the detector response is the bandwidth. A variety of different bandwidths between \SI{80}{MHz} and \SI{800}{MHz} are studied by applying a commonly available Butterworth high pass and low pass filter. All high-pass filters are modelled using a 5th order Butterworth filter, and all low-pass filters are modelled using a 10th order Butterworth filter. The bandwidth is specified by quoting the critical frequencies, i.e., the frequency where the filter response drops to \SI{-3}{dB}. In the following, we restrict ourselves to present only the results for an upper cutoff frequency of \SI{200}{MHz}, \SI{400}{MHz} and \SI{800}{MHz} to focus on the main general dependencies and because we found that additional changes of the lower cutoff frequency and/or further finetuning of the cutoff frequencies has little impact which we'll report at the end of Sec.~\ref{sec:sensitivity_LPDA}.

The simulation procedure is described as follows: For each bandwidth and trigger scheme, calculate the threshold that leads to the same trigger rate on thermal noise. Then use this threshold to compare the quantity of interest (e.g. the trigger turn-on curve for a given signal type (Sec.~\ref{sec:trigger}), or the sensitivity to neutrinos (Sec.~\ref{sec:sensitivity})). 

It was found that the bandwidth has a significant impact on the noise trigger rate. On the other hand, the dispersion of the signal chain does not influence the noise trigger rate as thermal noise has random phases. Thus, an additional shift in the phases does not change the properties of thermal noise.  Thermal noise is simulated in the frequency domain where the amplitude of each frequency bin is drawn from a Rayleigh distribution and the phase of each frequency bin is drawn from a uniform distribution \cite{NuRadioReco}. 

\begin{figure}[t]
    \centering
    \includegraphics[width=0.49\textwidth]{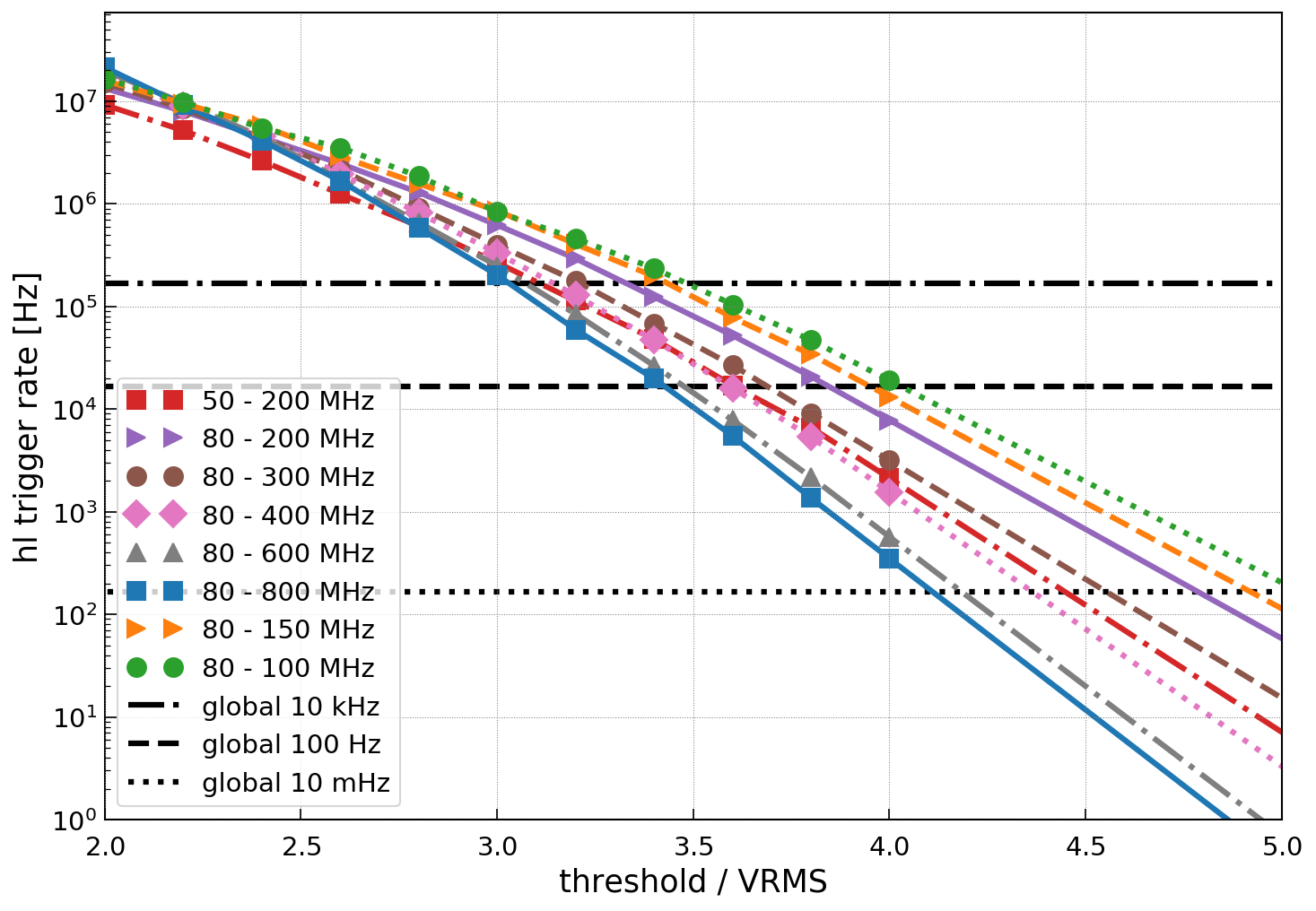}
    \includegraphics[width=0.49\textwidth]{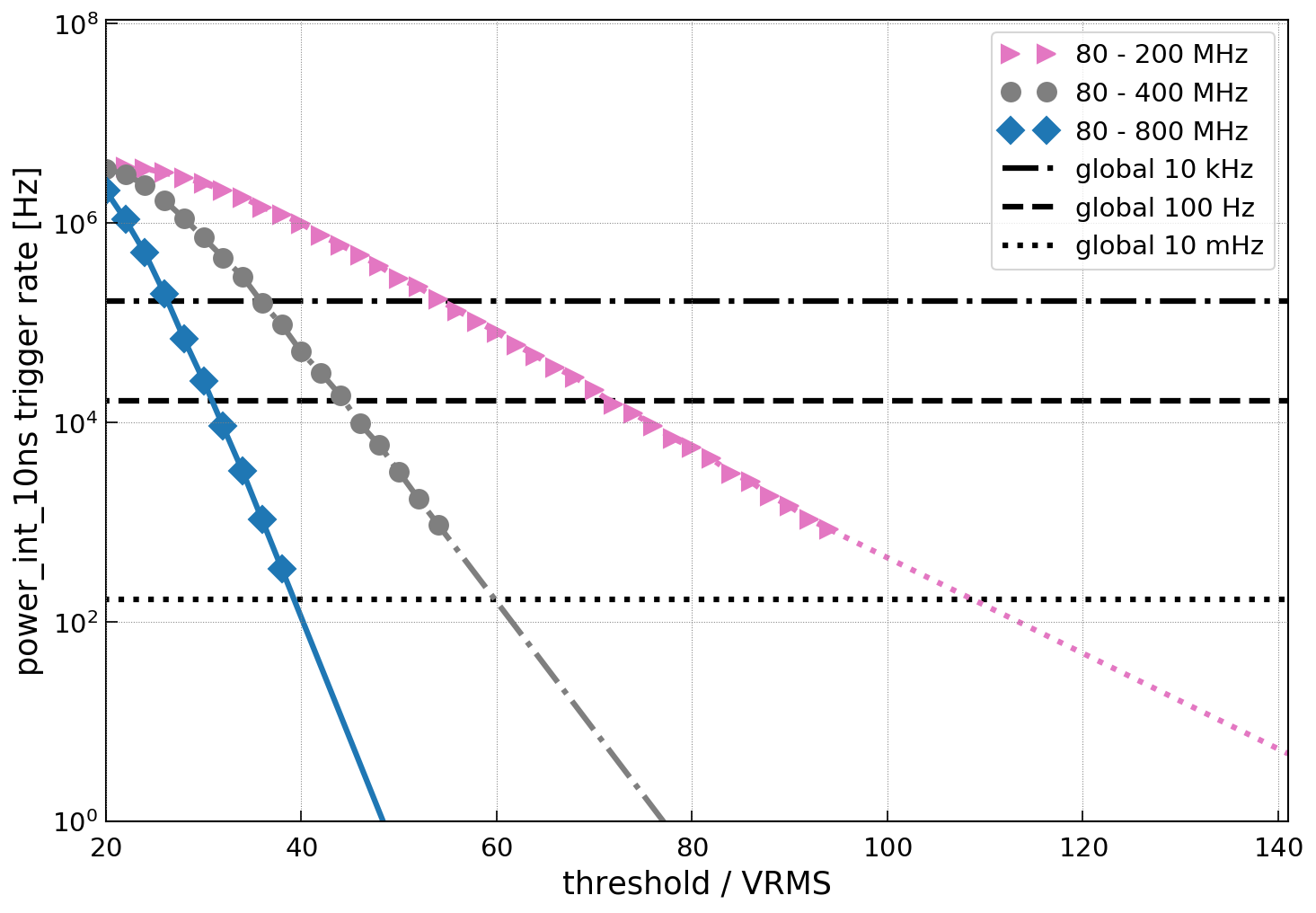}
    \caption{Single channel trigger rate as a function of threshold normalized to the noise RMS for a variety of different bandwidths as indicated by the legend. The lines show a linear interpolation/extrapolation to the simulated data points. Horizontal lines show the resulting global trigger rate for a 2 out of 4 majority logic with a coincidence time of \SI{30}{ns} between different channels. (left) high/low trigger. (right) power integration trigger with an integration window of \SI{10}{ns}.}
    \label{fig:trigger_rate}
\end{figure}

The resulting trigger rates as a function of threshold normalized to the RMS noise are shown in Fig.~\ref{fig:trigger_rate} for the high/low trigger (left panel) and a power integration trigger with an integration window of \SI{10}{ns} (right panel), for a variety of different bandwidths. 
A typical experimental setup is that one station consists of $N$ antennas and that a temporal coincidence between $N_c$ channels is required to further reduce the trigger rate. The coincidence window $\Delta t$ is adjusted to the maximum possible time delay between two antennas which is given by $\Delta t = d \times n / c$, where $d$ is the maximum distance between any pair of antennas, $c$ is the speed of light in vacuum and $n$ is the index of refraction of the medium surrounding the antennas. To be concrete, we consider a four antenna setup where at least two antennas need to have a trigger within a time window of \SI{30}{ns} is considered. This corresponds to the properties of the ARIANNA detector (see. e.g. \cite{ARIA, Anker_2020}). 
The global trigger rate $r_g$ is given by
\begin{equation}
    r_g = N_c {N\choose N_c} r_s^{N_c}  \Delta t ^ {N_c - 1} \, ,
\end{equation}
where $r_s$ is the single channel trigger rate.
The corresponding single-channel trigger rates for a global trigger rate of \SI{10}{kHz}, \SI{100}{Hz} and \SI{10}{mHz} are indicated with horizontal lines in Fig.~\ref{fig:trigger_rate}. We note that an ARA-like detector consists of more antennas with a larger separation. Thus, the values for $N$, $N_c$ and $\Delta t$ will change but the general behaviour that the trigger rate is reduced significantly by requiring a coincidence between antennas remains the same. 

For the high/low trigger, we observe the general trend that the trigger rate increases with decreasing bandwidth for the same threshold/RMS noise ratio. This is because an upward thermal noise fluctuation always has a small downward fluctuation on both sides because of the bandwidth limitation. This mexican-hat like waveform \cite{mexicanhat} gets more pronounced as the bandwidth is reduced. This results in a higher probability to produce an opposite sign fluctuation, which increases the high/low trigger rate. We will see later that the sensitivity to neutrinos is increased with a decreased bandwidth despite the higher trigger thresholds. 

The general behavior observed for the high/low trigger is replicated by the power integration trigger. This behavior can be understood by noting that the low frequency content corresponds to large wave periods $T = 1/f$. As we keep the integration window $\Delta t$ constant, a upward fluctuation in one frequency bin is less often compensated by a downward fluctuation in another frequency bin. 

The threshold vs. rate calculation is also performed for the other trigger settings listed above and used in the following section to determine the threshold that corresponds to a global trigger rate of \SI{100}{Hz}.

\section{Comparison of trigger schemes}
\label{sec:trigger}
In this section we answer the question: What is the optimal\footnote{We do not claim to be complete here. We study the trigger schemes that have been used so far by different experiments. We do not exclude that one can come up with a more efficient trigger logic.} trigger scheme for a given detector response.

\begin{figure}[tp]
    \centering
    \includegraphics[width=0.49\textwidth]{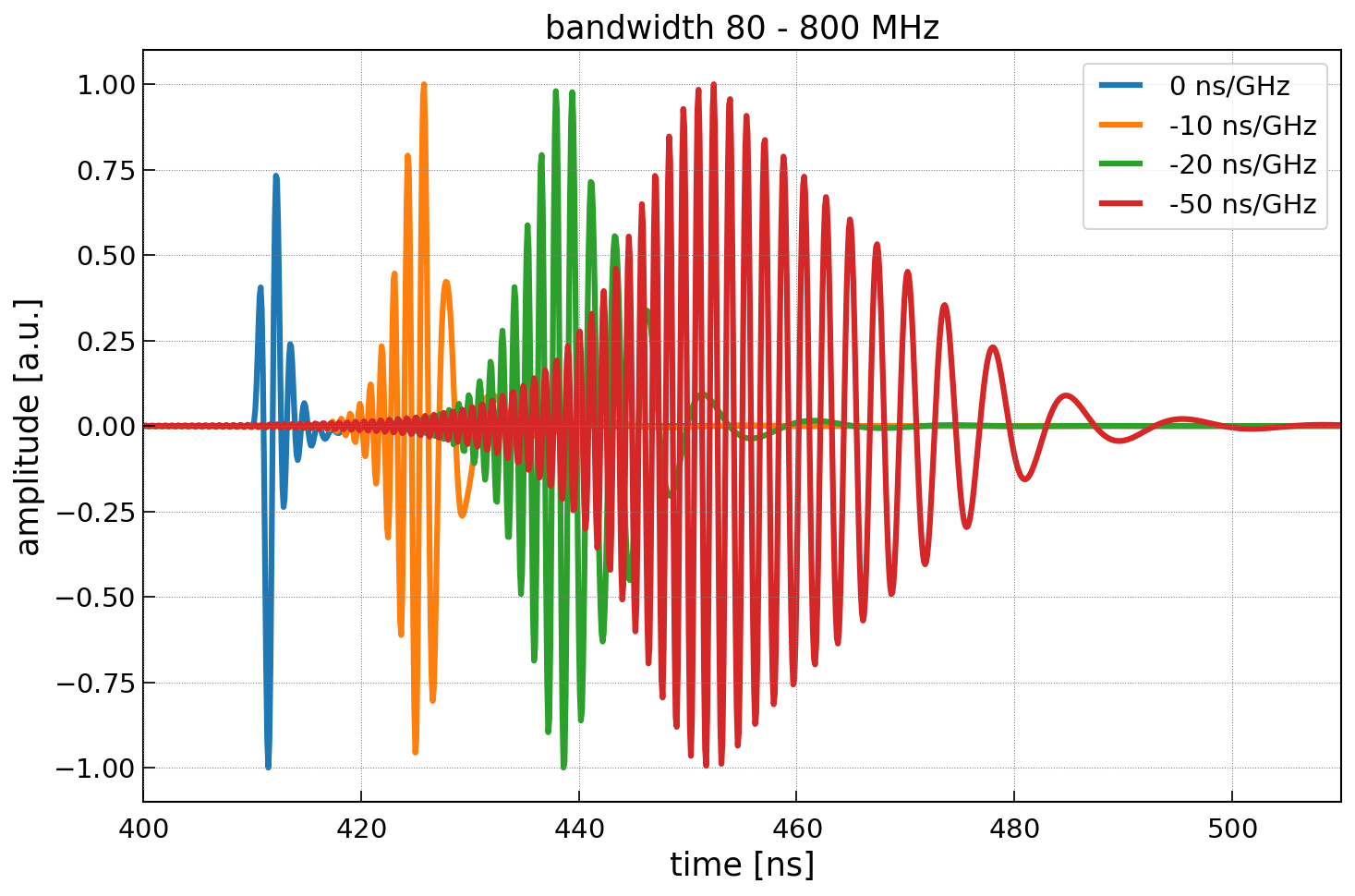}
    \includegraphics[width=0.49\textwidth]{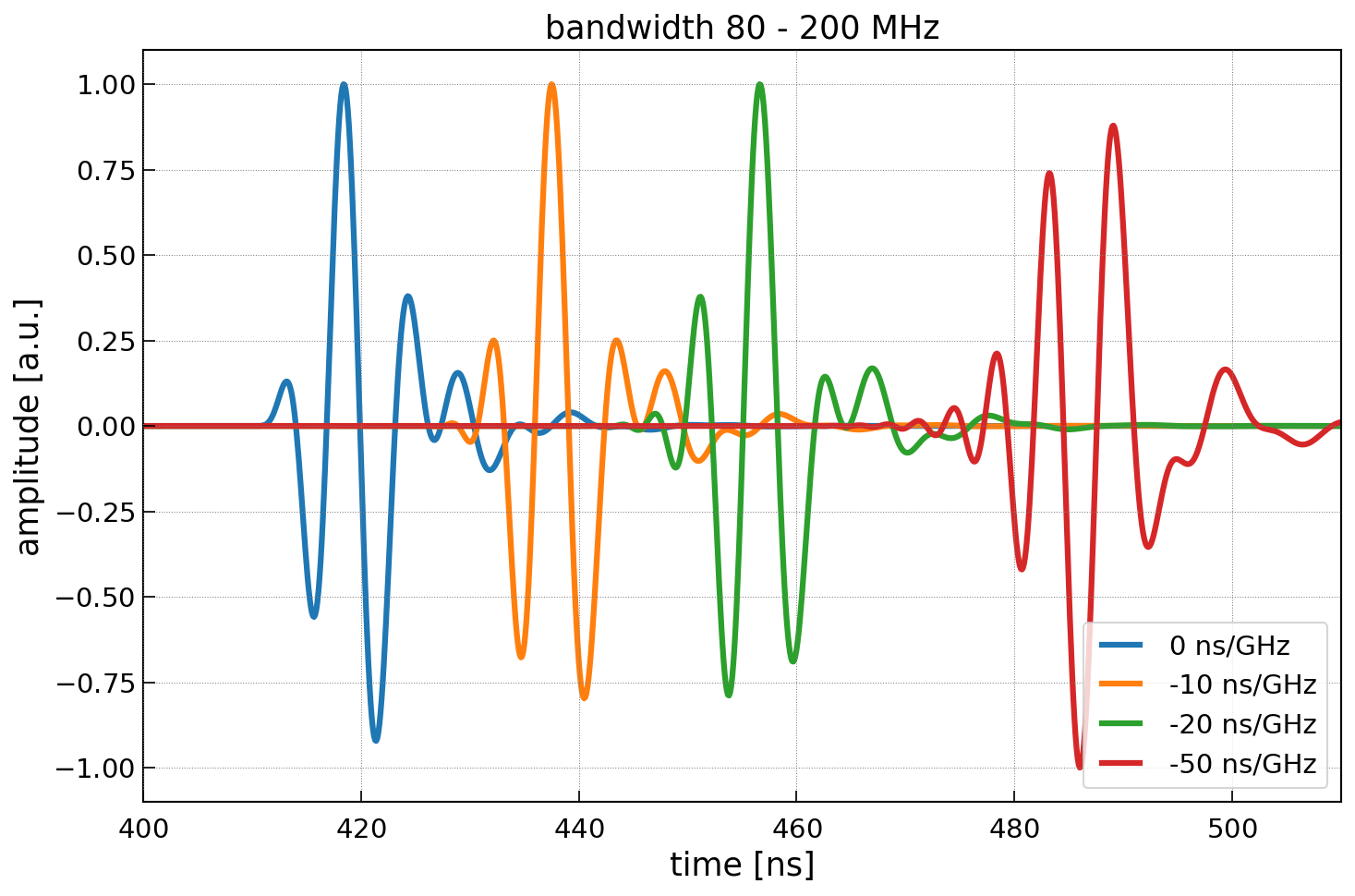}
    \caption{Example of bandwidth limited bipolar pulses for different linear dispersion as indicated by the legend. (left) For a bandwidth of \SI{80}{MHz} - \SI{800}{MHz}. (right) For a bandwidth of \SI{80}{MHz} - \SI{200}{MHz}.}
    \label{fig:pulse_example}
\end{figure}

In the first part of the section, generic waveforms are studied to derive general dependencies. The Askayan pulse is modelled with a bipolar delta pulse, i.e., with two succeeding samples in the time domain being 1 and -1 whereas the rest of the samples being zero. Then, the pulse is bandpass filtered and a linear group delay of the form $-X$~ns/GHz is applied. Examples of the resulting waveforms are shown in Fig.~\ref{fig:pulse_example} where the maximum amplitude is normalized to 1 and an arbitrary time offset is applied for better visibility. This procedure is motivated by the fact that the observed shape of the Askaryan waveforms is dominated by the detector response (see e.g. \cite{Anker_2020}). In the second part we study a more realistic signal calculation but we start with this toy study to gain general insights into the dependencies. 

A first relevant observation is that the dispersion has little impact on the resulting waveforms if the bandwidth is as small as \SI{80}{MHz} - \SI{200}{MHz}. This will have impact on the detector design. Normally the signal chain is designed to reduce dispersion as this would reduce the signal amplitude, but if a low trigger bandwidth is chosen, the dispersion has little effect and the waveform is mostly determined by the bandwidth. 

\begin{figure}[tp]
    \centering
    \includegraphics[width=0.49\textwidth]{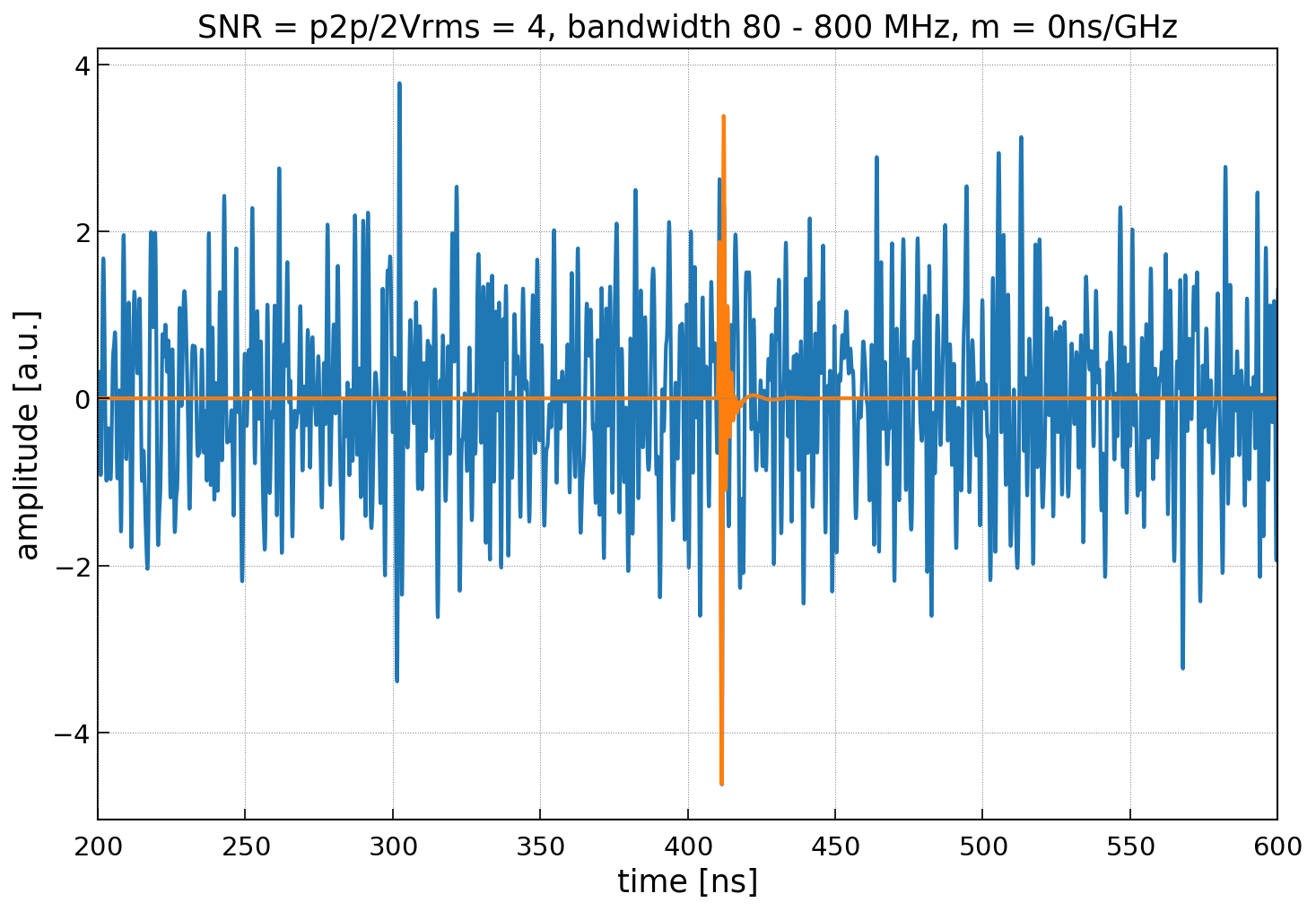}
    \includegraphics[width=0.49\textwidth]{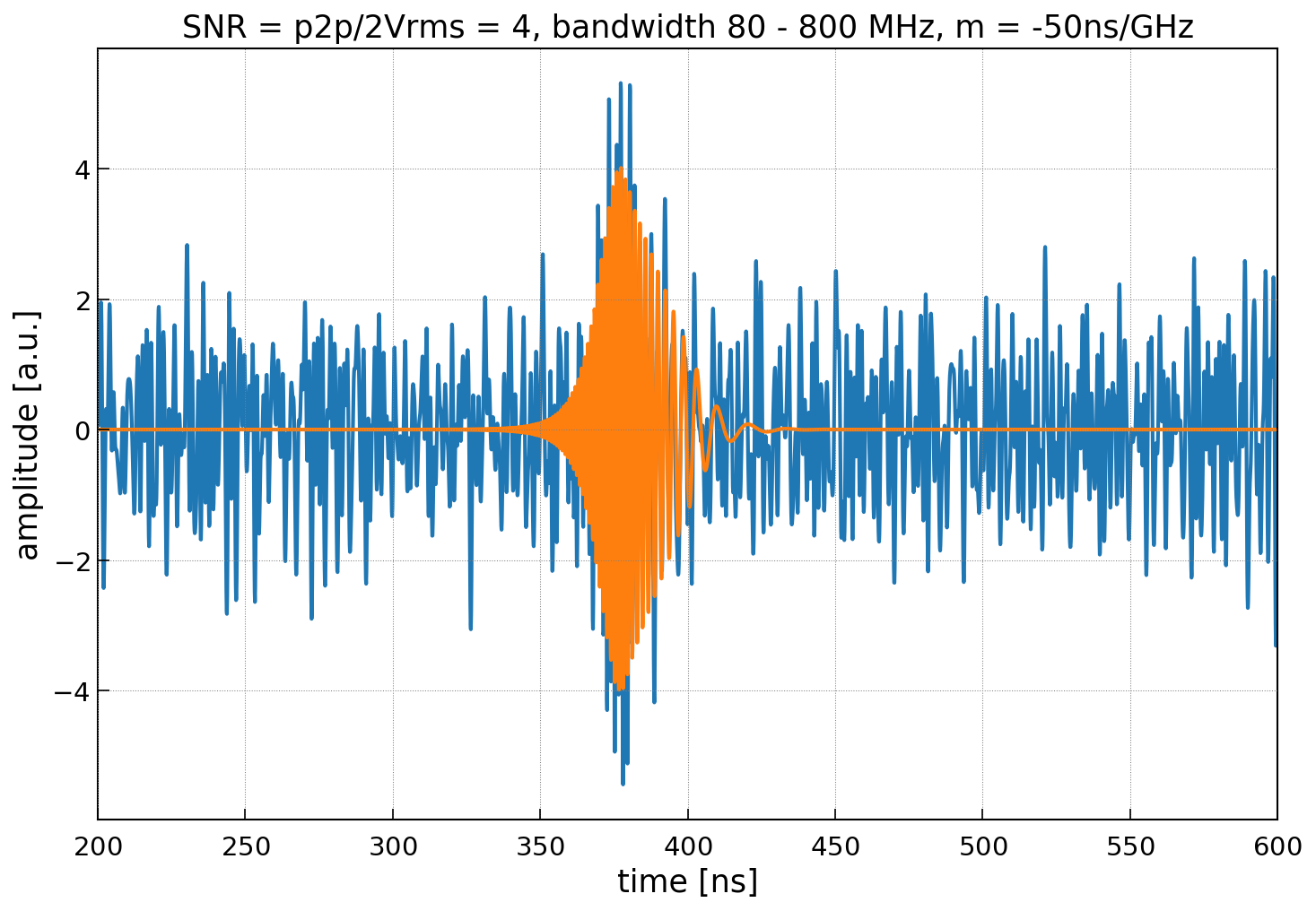}
    \includegraphics[width=0.49\textwidth]{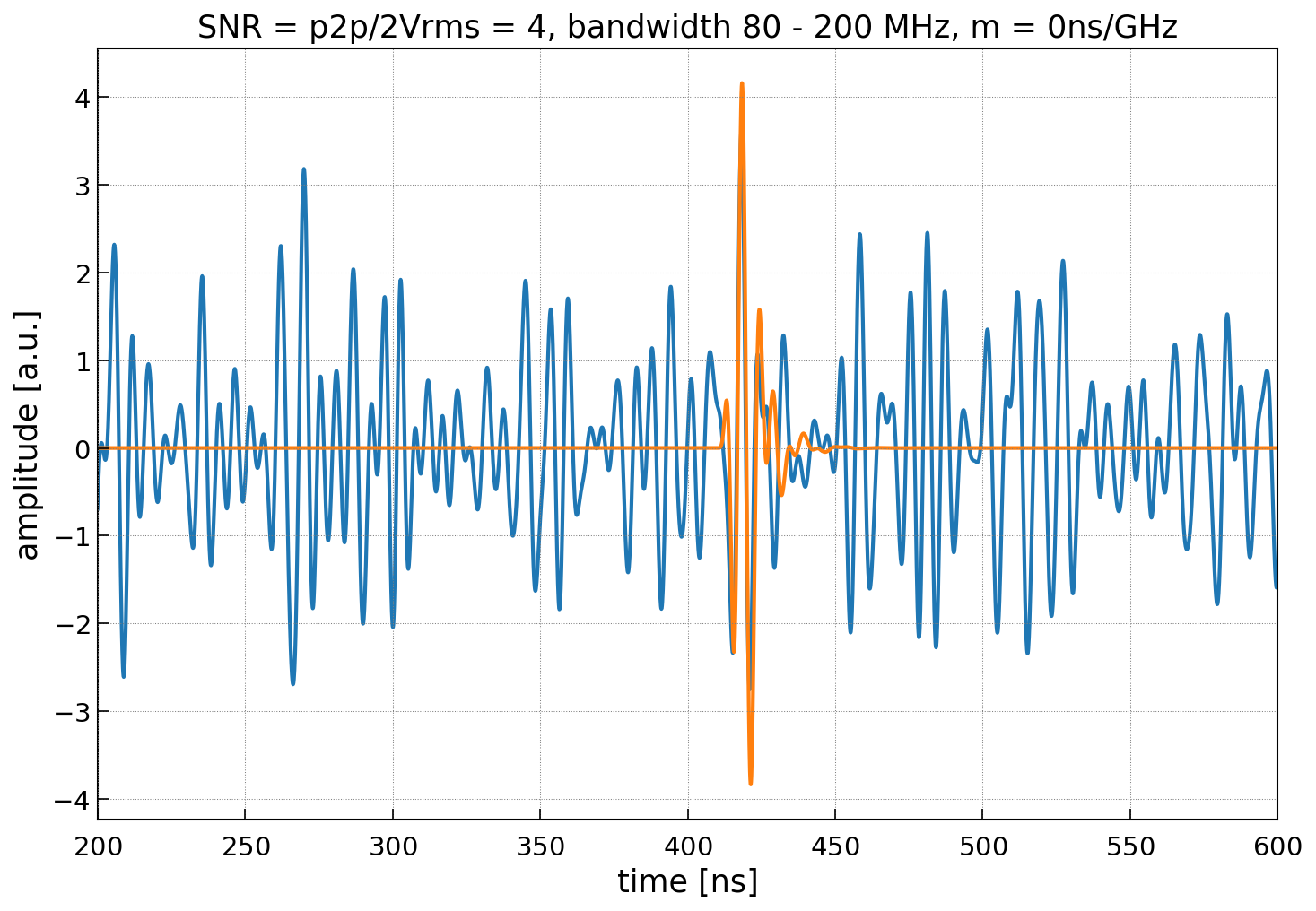}
    \includegraphics[width=0.49\textwidth]{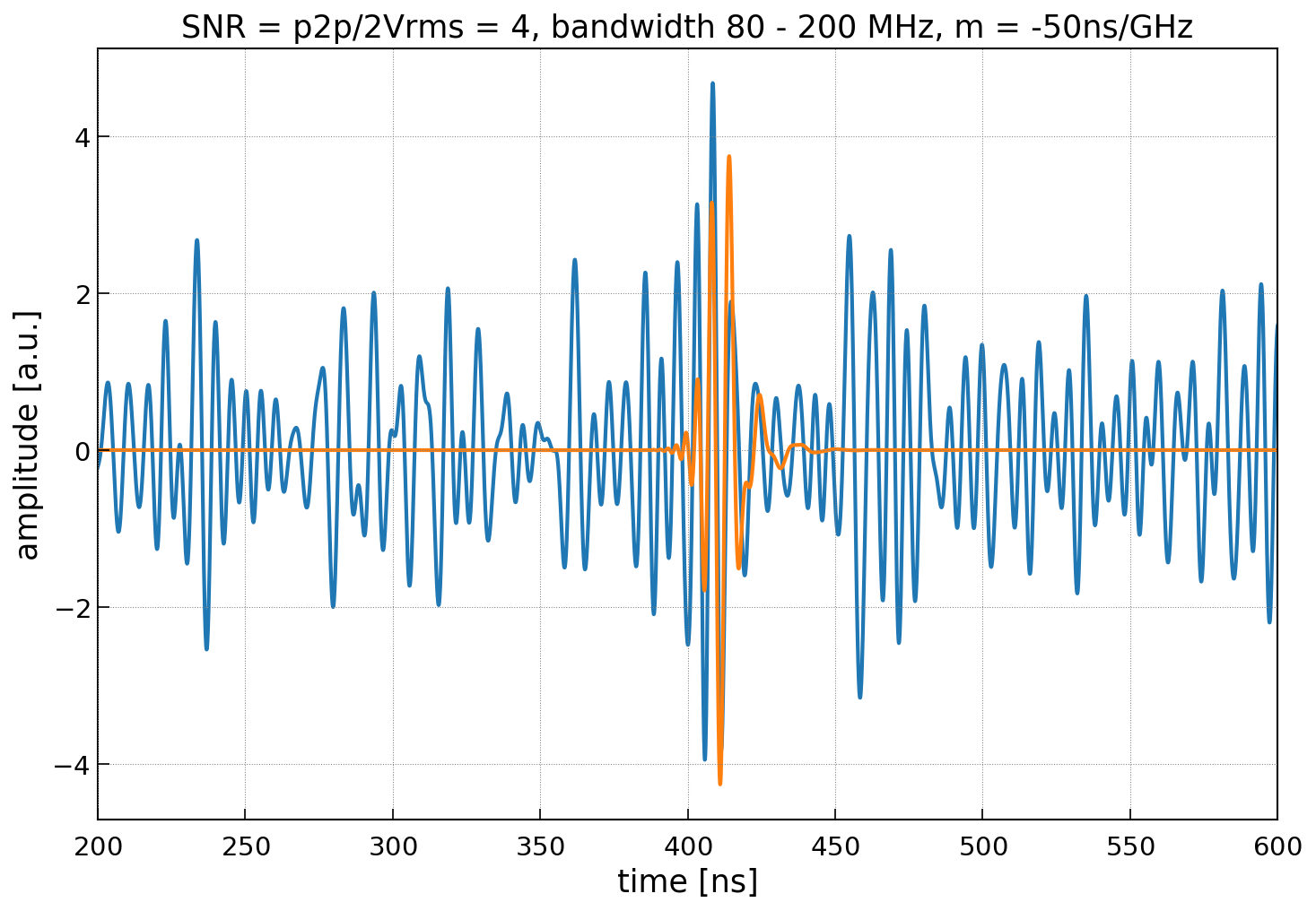}
    \caption{Example of bandwidth limited and dispersed bipolar pulses with noise. The maximum peak-to-peak amplitude of the (noiseless) signal is normalized to 8 times the RMS noise which is often referred to as a 4 sigma signal. The orange curves show the signal and the blue curves shows signal + noise. The upper panels are for a bandwidth of \SI{80}{MHz} - \SI{800}{MHz} and the lower panels for a bandwidth of \SI{80}{MHz} - \SI{200}{MHz}. The left panels are for no signal dispersion, whereas the right panels are for a signal dispersion of \SI{-50}{ns/GHz}.}
    \label{fig:pulse_example_noise}
\end{figure}

In Fig.~\ref{fig:pulse_example_noise} another example is presented that shows the signal pulses embedded in thermal noise. The noise is inserted before the bandpass filter so that the noise has the same frequency content as the filter response as observed experimentally. The signal pulses are normalized to have the same peak-to-peak amplitude. 

The trigger efficiency for each trigger scheme, bandwidth, dispersion, and for a global trigger rate of \SI{100}{Hz} is determined as follows. In our toy study, the bandwidth and dispersion determines the pulse form (cf. Fig.~\ref{fig:pulse_example}). For each possible waveform, i.e., combination of bandwidth and dispersion, the trigger efficiency is calculated for each trigger scheme. We scan through discrete value of the maximum signal amplitude ranging from 0 to about 10 times the RMS noise. For each choice of signal amplitude, 500 different noise realizations are generated (cf. Fig.~\ref{fig:pulse_example_noise}) and the trigger is evaluated. The trigger efficiency is then given by the number of cases that fulfills the trigger condition divided by the number of trials (500 in our case). 

At a thermal noise trigger rate of \SI{100}{Hz}, the probability to trigger on a noise fluctuation is negligible. Each trace has 512 samples with a time binning of \SI{0.5}{ns} resulting in a trace length of \SI{256}{ns}. The probability of getting a trigger from a pure noise fluctuation in any of the 500 trials is $\SI{100}{Hz} \times \SI{256}{ns} * 500 = 1.3\%$. Thus, the calculated trigger efficiency is not increased by more than $\sim1\%$ which is sufficiently low for the level of precision of a few percent we want to achieve here. 

\begin{figure}[tp]
    \centering
    \includegraphics[width=0.45\textwidth]{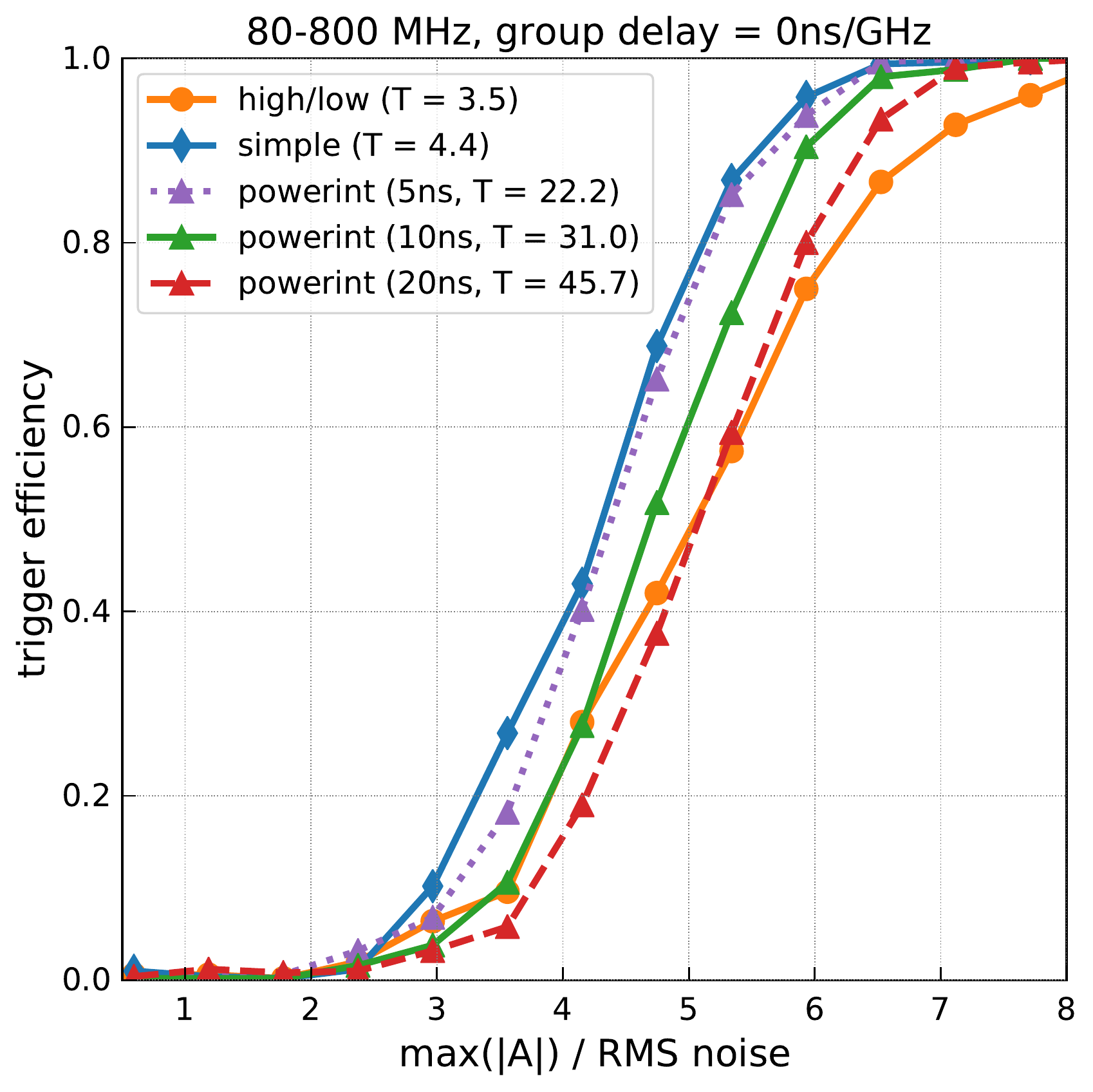}
        \includegraphics[width=0.45\textwidth]{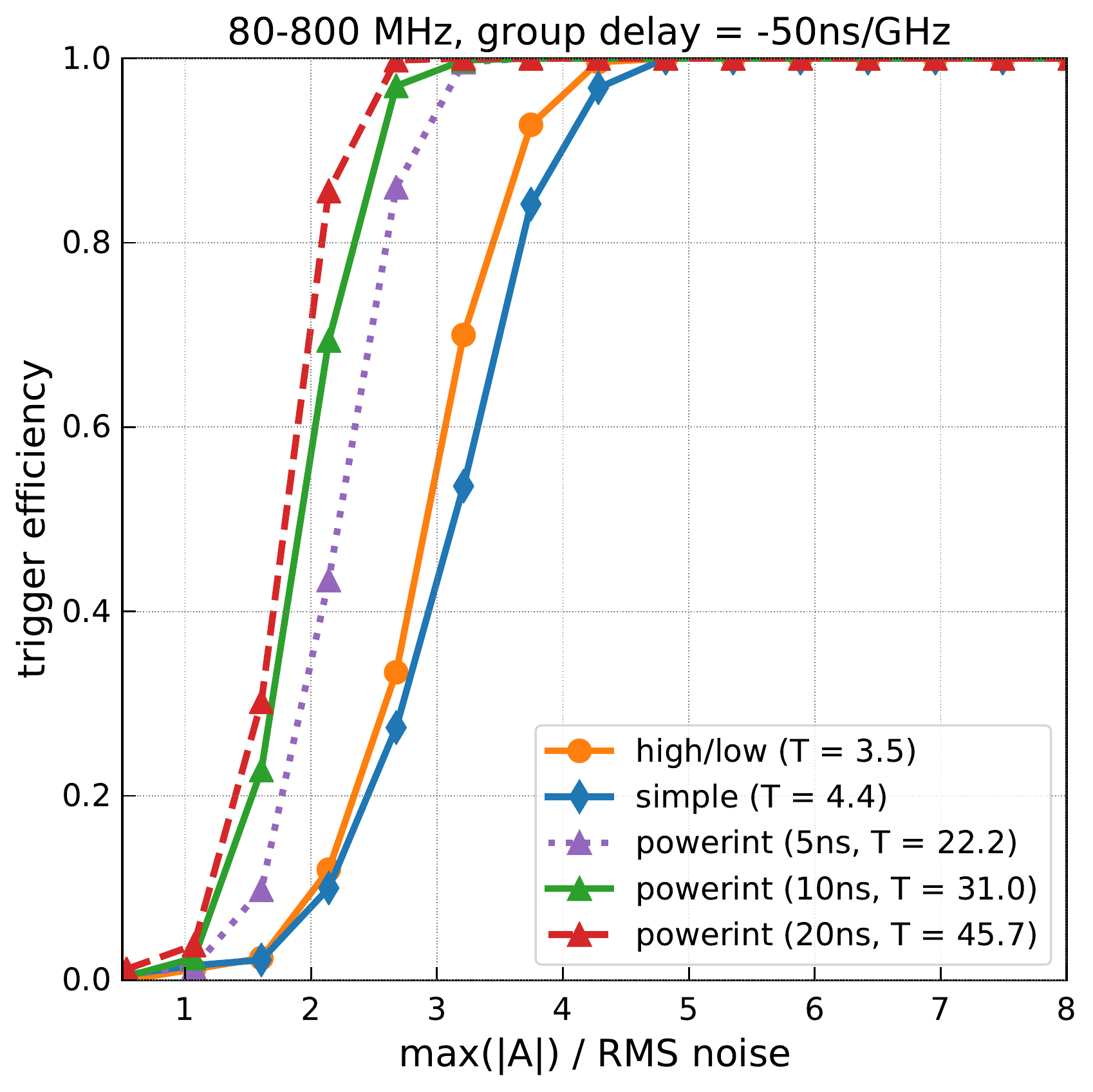}
    \caption{Trigger efficiency as a function of maximum signal amplitude normalized to the RMS noise for a bandwidth of \SI{80}{MHz} - \SI{800}{MHz}. The different curves correspond to the different trigger schemes as indicated by the legend. The legend also specifies the threshold for each trigger ($T = X$) that corresponds to the same thermal noise trigger rate of \SI{100}{Hz}. (left) For undispersed signals. (right) for a dispersion of \SI{-50}{ns/GHz}.}
    \label{fig:trigger_efficiency_800MHz}
\end{figure}

First, we discuss the results for a large bandwidth of \SI{80}{MHz} - \SI{800}{MHz} and the two extreme cases of no dispersion and a dispersion of \SI{-50}{ns/GHz}. We note that the group delay of the ARIANNA LPDA antennas varies within \SI{30}{ns} in this band. Thus, a dispersion of \SI{-50}{ns/GHz} approximates an LPDA setup with additional dispersion introduced by filters and amplifiers (cf. \cite{ARIANNATimeDomain}). More realistic responses will be studied later. 
The resulting trigger efficiencies are presented in Fig.~\ref{fig:trigger_efficiency_800MHz}. We observe that a power integration trigger, where the integration window is matched to the width of the signal, generally performs best. For the case of no dispersion, the power integration trigger with a \SI{5}{ns} integration time works better than larger integration windows. For the case of strong dispersion, the power integration trigger with \SI{20}{ns} integration window works best. Interestingly, the simple threshold trigger works slightly better than the best power integration trigger for the case of no signal dispersion, however for the case of a large signal dispersion it performs the worst. The high/low trigger performs always worse than the power integration trigger. For the no dispersion case we attribute this behaviour to the fact that the signal pulse (blue curve in Fig.~\ref{fig:pulse_example} left) is not symmetric. For the case of large dispersion case, the observed behaviour follows the intuition that integrating over the full length of the signal takes into account more information than just evaluating a high and low value. 

We note that Fig.~\ref{fig:trigger_efficiency_800MHz} does not imply that a dispersed signal leads to a better sensitivity. This is an artifact of plotting the trigger efficiency as a function of the maximum signal amplitude of the dispersed signal. For the same signal amplitude, the dispersed signal corresponds to a much larger fluence as directly visible from Fig.~\ref{fig:pulse_example}. Instead, this analysis answers the question: What is the optimal trigger for a given signal shape. And as the signal shape is often predominantly determined by the detector response, it can be generalized to the question: What is the optimal trigger for a given detector response.

\begin{figure}[tp]
    \centering
    \includegraphics[width=0.45\textwidth]{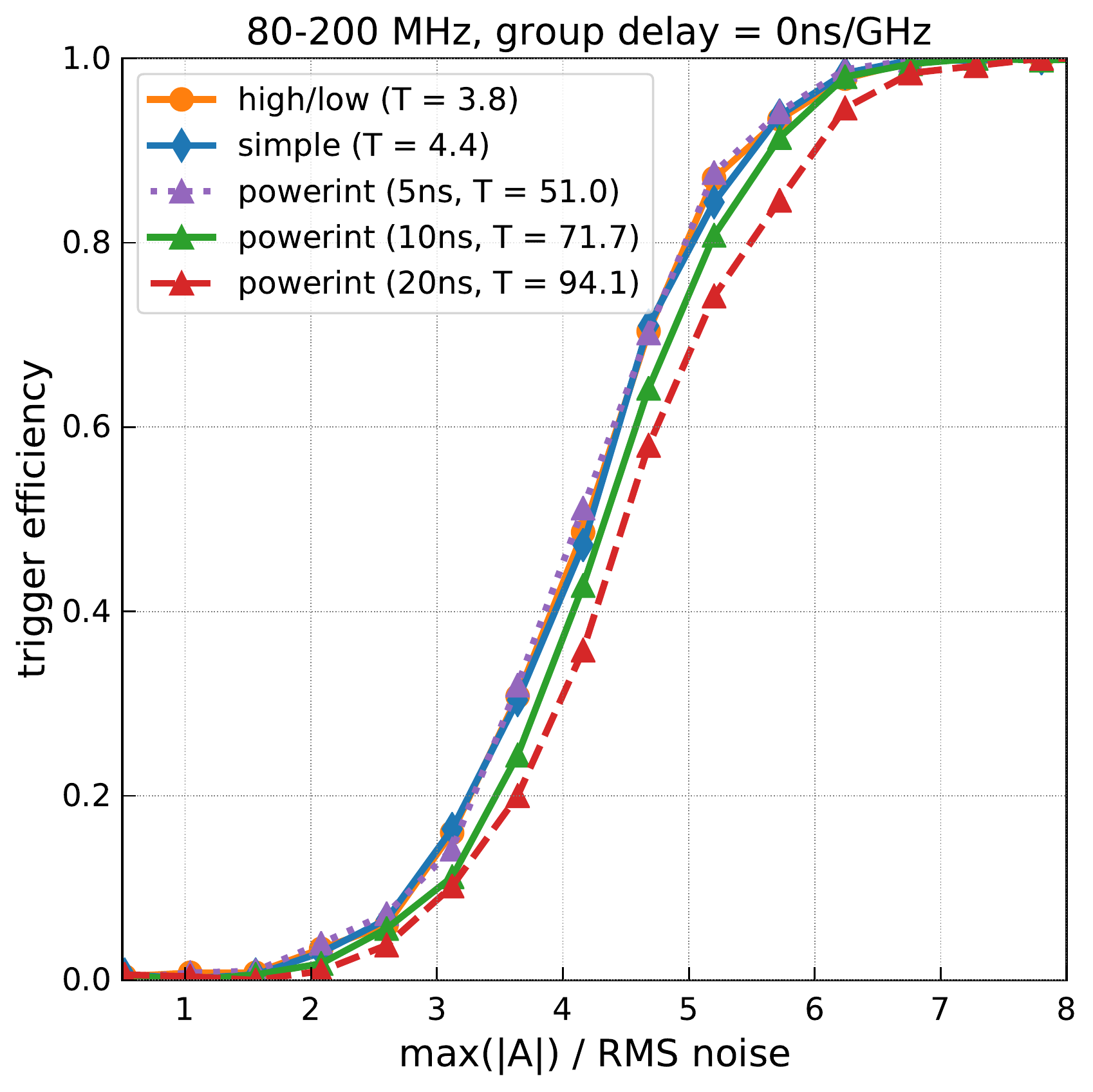}
    \includegraphics[width=0.45\textwidth]{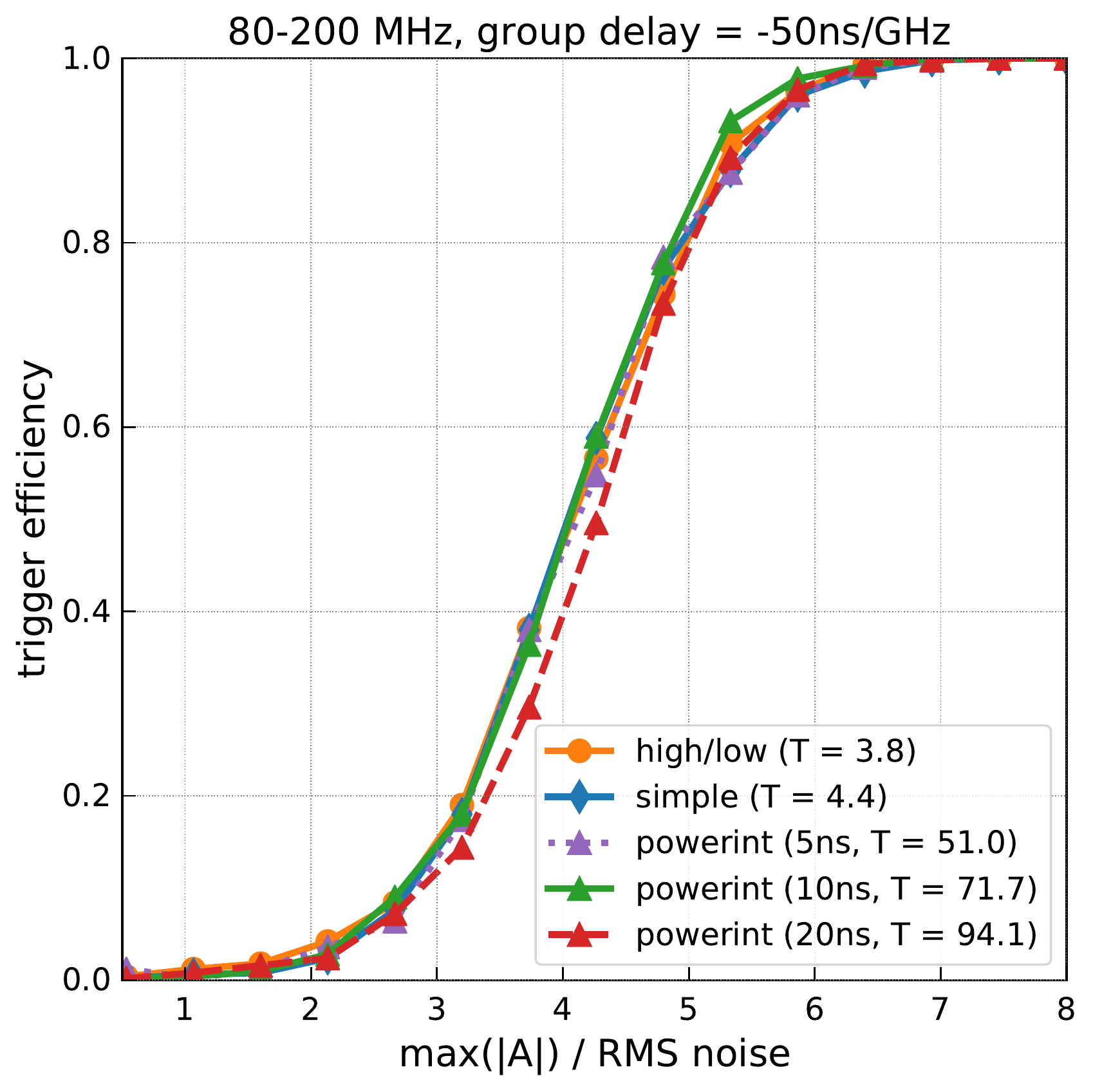}
    \caption{Same as Fig.~\ref{fig:trigger_efficiency_800MHz} but for a bandwidth of \SI{80}{MHz} - \SI{200}{MHz}.}
    \label{fig:trigger_efficiency_200MHz}
\end{figure}

The analysis is repeated for a bandwidth of  \SI{80}{MHz} - \SI{200}{MHz} and the results are presented in Fig.~\ref{fig:trigger_efficiency_200MHz}. We find that the choice of trigger is essentially irrelevant and all different trigger schemes lead to the same trigger efficiency curve. The only exception is the \SI{20}{ns} power integration trigger that performs slightly worse for the case of no signal dispersion. 

From this toy study we derive the following general behaviour of the trigger efficiency: A power integration trigger whose integration window is matched to the extent of the signal performs generally best. For the case of a small trigger bandwidth, the trigger efficiency is similar for all cases that were studied. 

The behaviour shown in Fig.~\ref{fig:trigger_efficiency_800MHz} and \ref{fig:trigger_efficiency_200MHz} also illustrates why the performance of two different detectors cannot be compared by comparing the trigger efficiency curves. A less dispersive detector will have in general a better performance although the trigger efficiency turns on at a higher threshold. 
Radio neutrino detectors utilize a variety of antenna types such as high-gain LPDAs in surface designs and lower gain bicones in boreholes. The only way to correctly compare the sensitivity of different detector designs is by simulating the number of observable neutrinos for a representative set of neutrino interactions in the ice which we will do in the next section to study the impact of bandpass filters. 

This analysis also shows that the translation of a power integration threshold to an equivalent amplitude threshold depends strongly on the dispersion of the detector. Thus, the specification of an equivalent single-antenna threshold of an interferometric phased array system, which uses a power integration trigger, needs to be interpreted with caution. In \cite{ARAprogress}, the equivalent amplitude threshold of a 7 channel phased array system was measured in-situ using a calibration pulser and yielded a single-antenna amplitude threshold of approx. two times the RMS noise for an integration window of \SI{10.7}{ns}. This result was often used to approximate a phased array system with a single antenna with an amplitude threshold of two times the RMS noise, e.g., in a MC simulation to estimate the sensitivity to neutrinos \cite{ARAprogress, Gen2WhitePaper}. This leads to potential errors as the input signal, dispersion and bandwidth of the simulated detectors are different. Thus, a full simulation of a phased array system when estimating the neutrino sensitivity is advisable.

\subsection{Realistic signal prediction}
\begin{figure}[tp]
    \centering
    \includegraphics[width=0.32\textwidth]{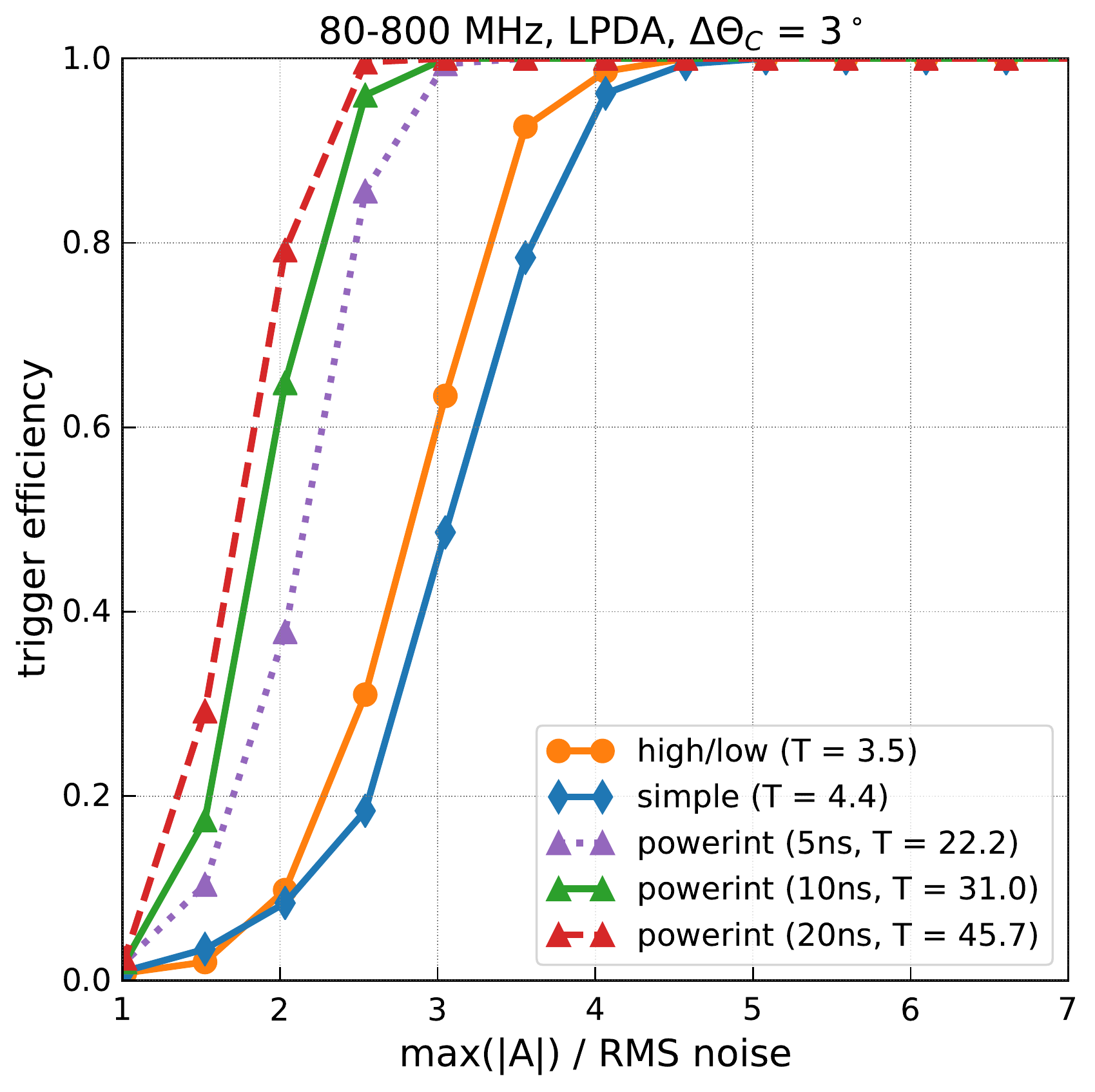}
    \includegraphics[width=0.32\textwidth]{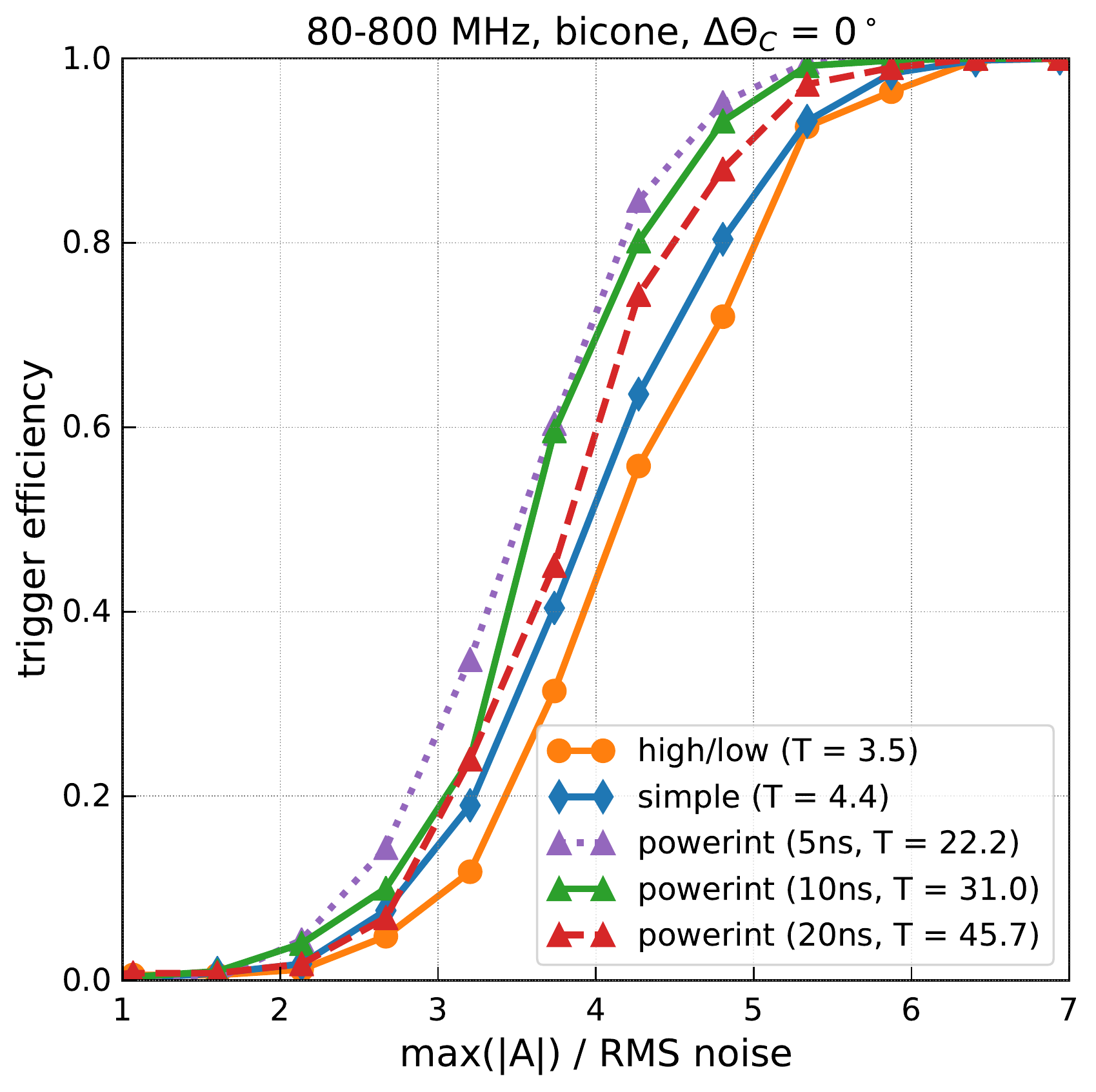}
    \includegraphics[width=0.32\textwidth]{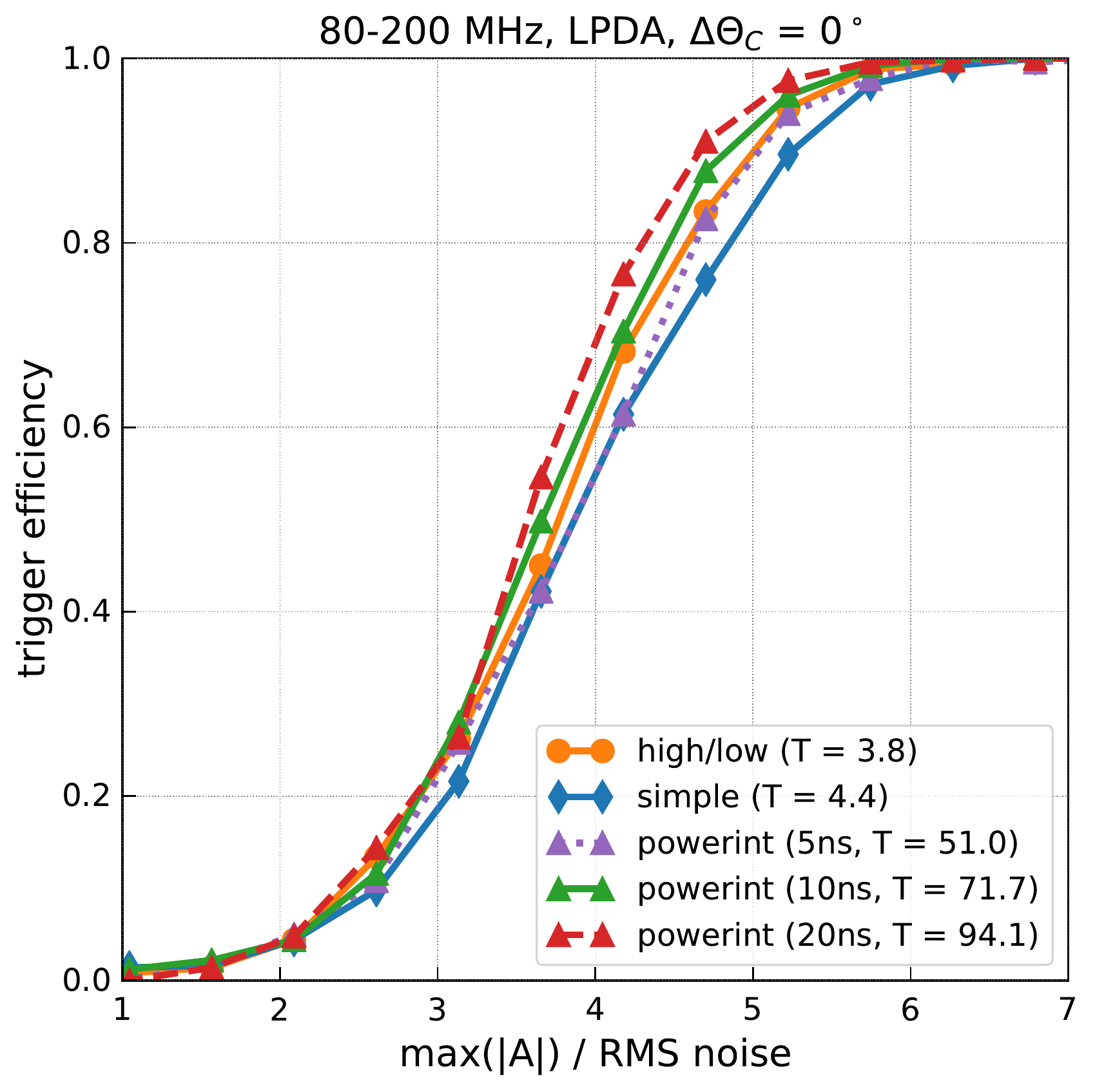}
    \caption{Same as Fig.~\ref{fig:trigger_efficiency_800MHz} but for a realistic Askaryan signal expectation. (left) \SI{3}{\degree} off-cone Askaryan signal convolved with an LPDA antenna response and a \SI{80}{MHz} to \SI{800}{MHz} bandpass filter. (middle) on-cone Askaryan signal convolved with a bicone antenna response and a \SI{80}{MHz} to \SI{800}{MHz} bandpass filter. (right) on-cone Askaryan signal convolved with an LPDA antenna response and a \SI{80}{MHz} to \SI{200}{MHz} bandpass filter.}
    \label{fig:trigger_efficiency_LPDA_bicone}
\end{figure}

In the following, we repeat the analysis for a more realistic signal prediction that will confirm the conclusions of the toy study above. The Askaryan signal is calculated using the \emph{Alvarez2000} model \cite{AlvarezMuiz2000, NuRadioMC}, which is a parameterization of a microscopic simulation of the in-ice shower that follows a neutrino interaction and its radio emission using the ZHS code \cite{ZHS}.  We calculate the Askaryan signal once for an observer at the Cherenkov angle and once for an observer \SI{3}{\degree} away from the Cherenkov angle which corresponds to typically expected signals. The Askaryan signal is convolved with an antenna response. We tested the two typical antenna types used for neutrino detection. A LPDA antenna as used for the surface antennas by the ARIANNA and RNO-G detectors \cite{ARIANNATimeDomain}, and a bicone antenna as used for deep antennas by the ARA and RNO-G detectors. The main difference between these two antennas that is relevant for this study is that the LPDA is  more dispersive than the bicone antenna. To complete the signal chain, the signal is bandpass filtered using the same filter as in the generic study above. The resulting trigger efficiencies of three of these scenarios are shown in Fig.~\ref{fig:trigger_efficiency_LPDA_bicone}.

Approximating the full signal chain with just a Butterworth bandpass filter corresponds to an ideal case as little filter-related dispersion is introduced. In reality, the dispersion of the signal chain is often larger. Thus, we also replaced the generic filter with the measured filter and amplifier response of the ARIANNA-HRA detector \cite{Anker_2020}. 

For all of these more realistic scenarios we find the same general trend: A power integration trigger where the integration window is matched to the signal width performs best or equally well than a different trigger scheme. The differences between the different triggers schemes are largest for extended signals, i.e., for a large bandwidth and large dispersion. If the trigger bandwidth is limited to \SI{80}{MHz} - \SI{200}{MHz}, the performance of all trigger schemes becomes relatively similar but with a slight preference to a power integration trigger. 

We note that these findings are also applicable for a interferometric phased array trigger which generates a synthesized waveform with the same signal shape but an increased signal-to-noise ratio. Thus, the same conclusions apply to this synthesized waveform.

\section{Effect of trigger bandwidth on sensitivity}
\label{sec:sensitivity}
In this section we answer the question: What is the impact of bandwidth on the neutrino sensitivity for a given trigger scheme and detector response. Current experiments such as ARA and ARIANNA use a large bandwidth of \SI{80}{MHz} to about \SI{1}{GHz} which is motivated by the prediction that the Askaryan signal extends up to GHz frequencies if observed on the Cherenkov cone. The sensitivity of the detector is quantiﬁed in terms of effective volume to an isotropic neutrino ﬂux. It is given by the weighted sum of all triggered events divided by the total number of events multiplied by the simulation volume and the simulated solid angle. The weighting factor is the probability of a neutrino reaching the simulation volume (and not being absorbed by the Earth). The effective volume is directly proportional to the number of observed neutrinos. 

\subsection{Simulation settings}
We perform this study for the proposed ARIANNA-200 detector, an array of 200 autonomous detector stations on the Ross Ice Shelf which uses two orthogonal pairs of downward pointing LPDA antennas to measure and trigger on Askaryan signals \cite{ARIANNA200}. An additional bicone antenna (as used by the ARA detector) at a depth of \SI{10}{m} is also part of the station to aid event reconstruction \cite{DnR2019} but not used for the trigger. 

The sensitivity to neutrinos for each trigger bandwidth and trigger scheme is simulated using NuRadioMC \cite{NuRadioMC} that uses NuRadioReco for the detector simulation \cite{NuRadioReco}.
We perform the simulation with the ice properties of the Moore's Bay site on the Ross Ice Shelf. We use the Alvarez2000 Askaryan emission model and assume a noise temperature of \SI{250}{K}. We assume a 1:1:1 flavor ratio between electron, muon and tau neutrinos, distribute arrival directions isotropically, and place the neutrino interactions uniformly in the ice around the detector.
The neutrino sensitivity is simulated at discrete neutrino energies and quantified via the effective volume per neutrino energy.

We observe that three parameters have a visible effect on the predicted neutrino sensitivity: The bandwidth of the detector, the trigger scheme and if we add thermal noise to the Askaryan signals or not in the simulation. Adding noise to the Askaryan signals increased the trigger probability on average, and it is thus important to include, but comes with the technical challenge to not count thermal noise fluctuation as detected neutrinos. We document the procedure to prevent that in appendix \ref{sec:noise}. In the following, all simulations are performed including thermal noise.

\subsection{Impact of bandwidth using the high/low trigger scheme}
\label{sec:sensitivity_LPDA}
We first discuss the effect of bandwidth for the high/low trigger scheme which is currently implemented in the ARIANNA electronics. 
According to the procedure outlined in Sec.~\ref{sec:method}, we first calculate the threshold that corresponds to a global thermal noise trigger rate of \SI{100}{Hz} (cf. Fig.~\ref{fig:trigger_rate} left). We find a threshold of approximately 3.4 times the RMS noise for the largest studied bandwidth of \SI{80}{MHz} - \SI{800}{MHz}. The trigger threshold increases for decreasing bandwidth. For the smallest bandwidth of \SI{80}{MHz} - \SI{200}{MHz} we obtain a trigger threshold of approximately 4 times the RMS noise. 
At the end of the section, the study will be repeated using a power integration trigger.

\begin{figure}[t]
    \centering
    \includegraphics[width=0.7\textwidth]{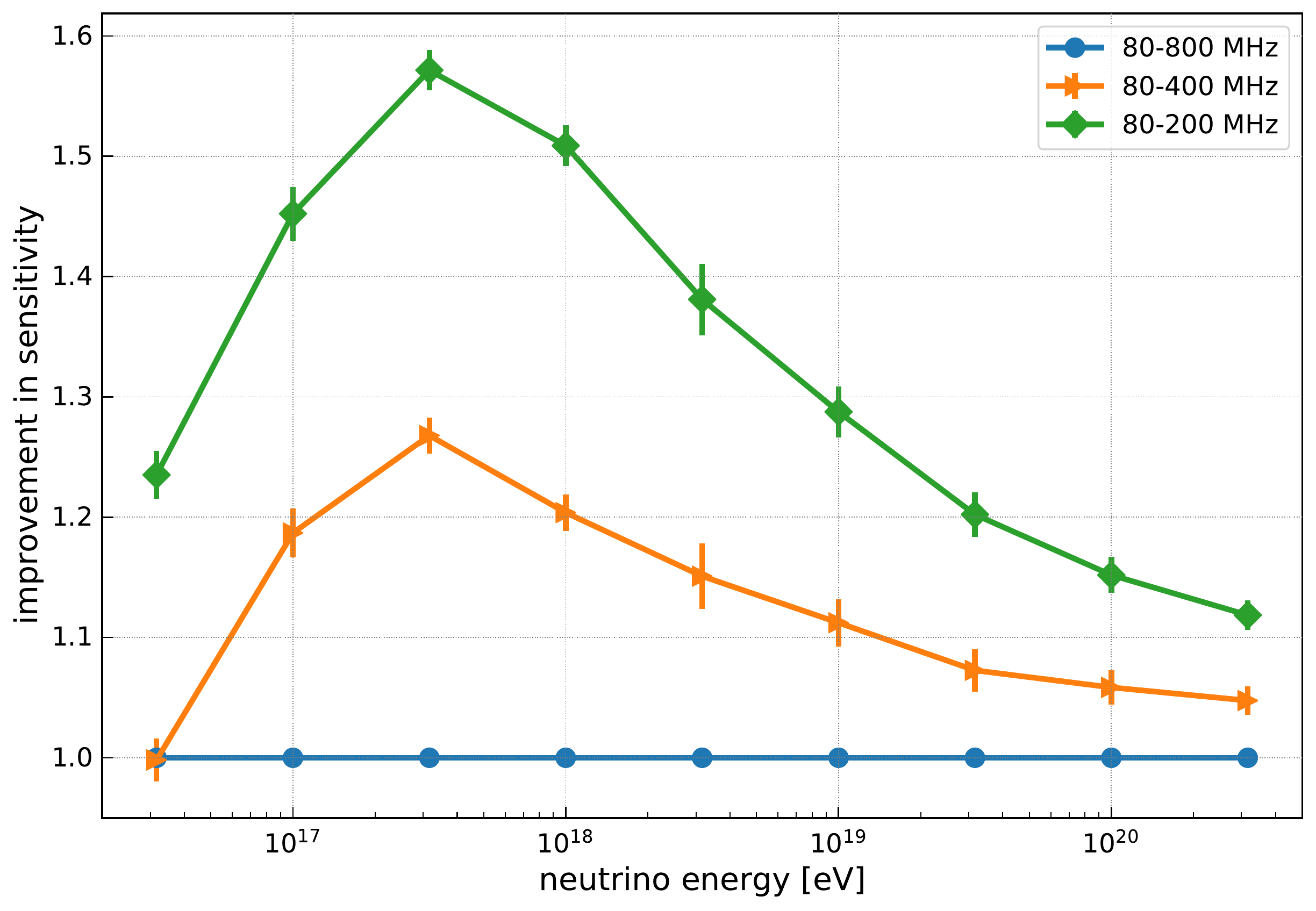}
    \caption{Impact of bandwidth on neutrino efficiency using the high/low trigger. The y-axis shows the improvement in sensitivity with respect to the largest studied bandwidth of \SI{80}{MHz} - \SI{800}{MHz}.}
    \label{fig:sensitivity_bandwidth_LPDA}
\end{figure}

In Fig.~\ref{fig:sensitivity_bandwidth_LPDA}, the increase in effective volume of a detector with a trigger bandwidth of \SI{80}{MHz} - \SI{200}{MHz} and \SI{80}{MHz} - \SI{400}{MHz} with respect to the reference with a large bandwidth of \SI{80}{MHz} - \SI{800}{MHz} is shown. We find that the sensitivity of an Askaryan detector can be increased by 50\% for neutrino energies between \SI{e17}{eV} and \SI{e18}{eV} with a simple bandpass filter despite running at a higher threshold. This finding illustrates that the trigger threshold alone does not give an adequate description of the performance of a radio neutrino detector. The bandwidth of the trigger circuit must be considered as well. We find that a reduced bandwidth of the trigger circuit strongly improves the sensitivity of the detector.

At first, this conclusion may seem counterintuitive because if a neutrino interaction is observed directly on the Cherenkov cone, the frequency content increases up to frequencies of \SI{1}{GHz}. To first order, the signal amplitude increases linearly with frequency up to a cutoff frequency after which the amplitude decreases quickly. 
Thus, for neutrino interactions observed on the Cherenkov cone, a larger bandwidth detects more power which should result in better sensitivity.

However, there are several more relevant factors working against that. First, with increasing angular distance between the observation angle and the Cherenkov angle, the cutoff frequency drops quickly into the hundred MHz range (see e.g. discussion in \cite{AlvarezMuiz2000, NuRadioMC}). For a given amplitude at the detector, the geometry of off-cone events is more favorable and leads to a greater number of observed events. 

Second, the effective area of radio antennas typically scale with the square of the wavelength of the incoming signal, leading to more detected power at lower frequencies. The relevant quantity is the vector effective length which is the proportionality factor between the incident electric field and the voltage output of the antenna. The vector effective length as function of frequency $\mathcal{H}(f)$ is related to the gain $G(f)$ as (see appendix of \cite{NuRadioReco} for details)
\begin{equation}
    \mathcal{H}(f) \propto \frac{1}{f} \sqrt{G(f)} \, .
\end{equation}
Thus, higher frequencies are naturally suppressed via $1/f$ and the antennas used in current Askaryan detectors like ARA and ARIANNA have the peak response between \SI{100}{MHz} and \SI{200}{MHz}.

Third, the RMS noise is typically flat in frequency. Then, the RMS noise is proportionally to the square root of the bandwidth. Thus, reducing the bandwidth will substantially reduce the RMS noise. For example reducing the bandwidth from \SI{720}{MHz} (\SI{80}{MHz} - \SI{800}{MHz}) to  \SI{120}{MHz} (\SI{80}{MHz} - \SI{200}{MHz}) reduces the RMS noise by a factor 2.5. 

\begin{figure}[t]
    \centering
    \includegraphics[width=0.6\textwidth]{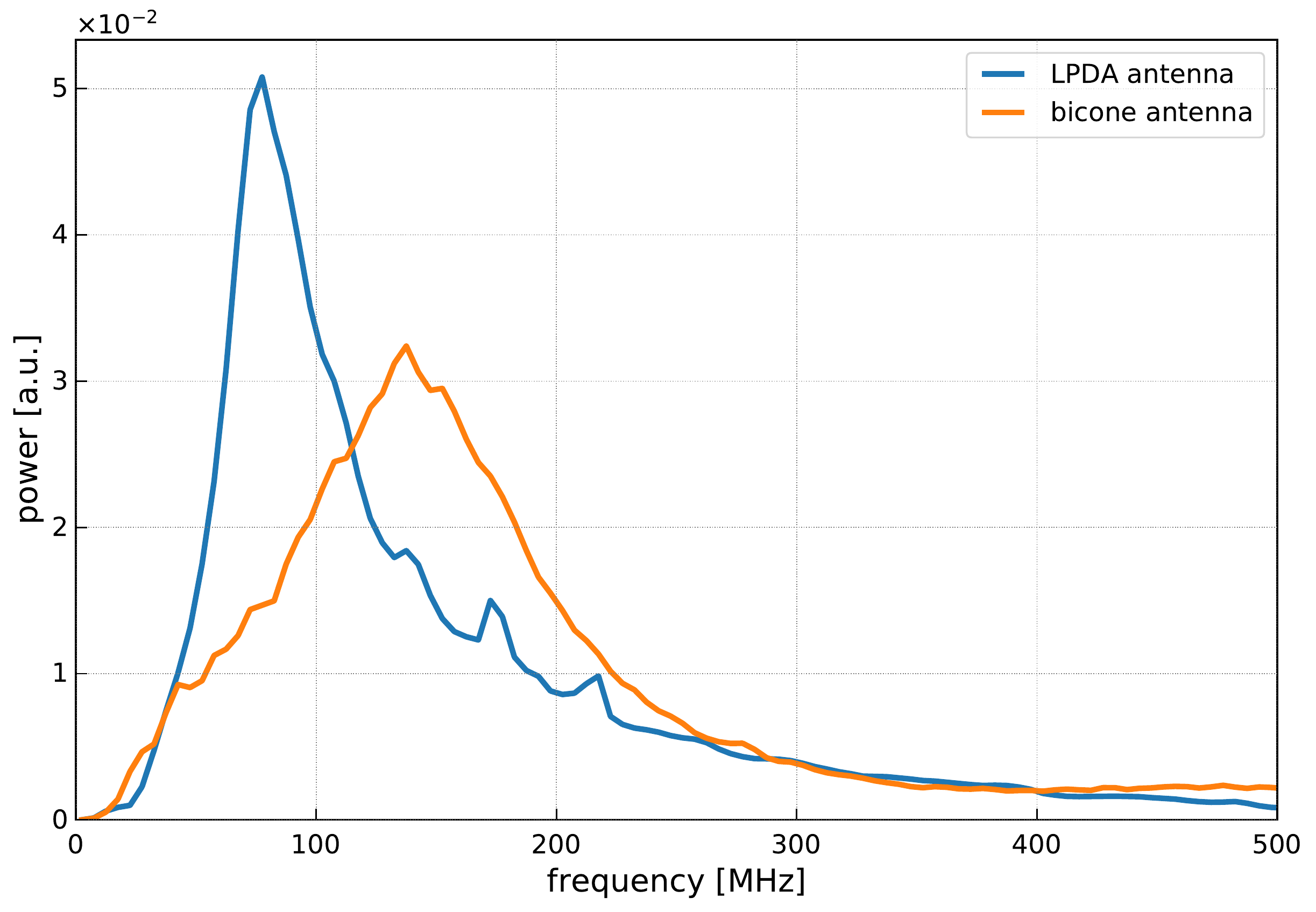}
    \caption{Averaged measured frequency spectrum of simulated neutrinos with an energy of \SI{e17}{eV}. The blue curve shows the average frequency spectrum recorded with an LPDA antenna and the orange curve shows the average frequency spectrum recorded with a bicone antenna for the same events.}
    \label{fig:freq_spetrum}
\end{figure}

To visualize these effects, we calculated the average frequency spectrum of triggered events which is shown in Fig.~\ref{fig:freq_spetrum}. For all simulated neutrinos with an energy of \SI{e17}{eV} that pass the high/low trigger with a bandwidth of \SI{80}{MHz} to \SI{800}{MHz}, we took the frequency spectrum after the antenna response was applied to the simulated Askaryan signals, normalized the power spectrum to the same integral, and summed up the contributions of all events. We observe that the resulting frequency spectrum is dominated by low frequencies. In case of an LDPA antenna, the spectrum is peaked at \SI{80}{MHz} which corresponds to the peak sensitivity of the LDPA antenna. We also analyzed the frequency spectrum measured by a bicone antenna where the peak in the frequency spectrum is at around \SI{150}{MHz}. This is close to the peak sensitivity of \SI{180}{MHz} of the bicone antenna (cf.~\cite{NuRadioReco}). In both cases, most signal power is concentrated at low frequencies. 

In summary, a narrow bandwidth trigger will reduce both the collected power of the Askaryan signal and the thermal noise. Due to the concentration of signal at the lower frequencies, especially for geometries off the Cherenkov cone, the reduction in noise power is greater than the reduction of signal power. Consequently, the sensitivity of the detector increases with decreasing bandwidth of the trigger. The effect is most pronounced at intermediate neutrino energies where many events are detected just above the detection threshold. At higher neutrino energies the signal amplitudes are often higher than the threshold which makes the high energies less dependent on changes of the trigger threshold. The weaker improvement at $10^{16.5}$~eV is more difficult to explain. A likely cause is that at these low energies, only showers that are observed directly on the Cherenkov cone produce a large-enough signal where the benefit of a reduced bandwidth is smaller. 

We studied a further fine tuning of the optimal bandwidth by e.g. decreasing the lower cutoff frequency to \SI{50}{MHz}, and by further variations of the upper cutoff frequency (e.g. to \SI{150}{MHz}) but did not find a significant additional gain in sensitivity. This is because an additional (small) variation in available bandwidth will not change the noise level significantly, compared to the factor of 2.5 reduction that is achieved by reducing the upper cutoff frequency from \SI{800}{MHz} to \SI{200}{MHz}. In practice, also RFI noise sources at lower frequencies discourage lowering the low frequency cutoff further.

\subsection{Impact of bandwidth using the power integration scheme}
We repeat the analysis using the power integration trigger scheme with integration windows of \SI{5}{ns}, \SI{10}{ns} and \SI{20}{ns}. Also for the power integration trigger we find that restricting the bandwidth leads to a larger neutrino sensitivity. Because the power integration trigger performs better than the high/low trigger at large bandwidth (cf. Fig.~\ref{fig:trigger_efficiency_LPDA_bicone}), the improvements in sensitivity are smaller and range between 30\% for the \SI{5}{ns} integration window to 20\% for the \SI{20}{ns} integration window at \SI{e18}{eV}.

\begin{figure}[t]
    \centering
    \includegraphics[width=0.7\textwidth]{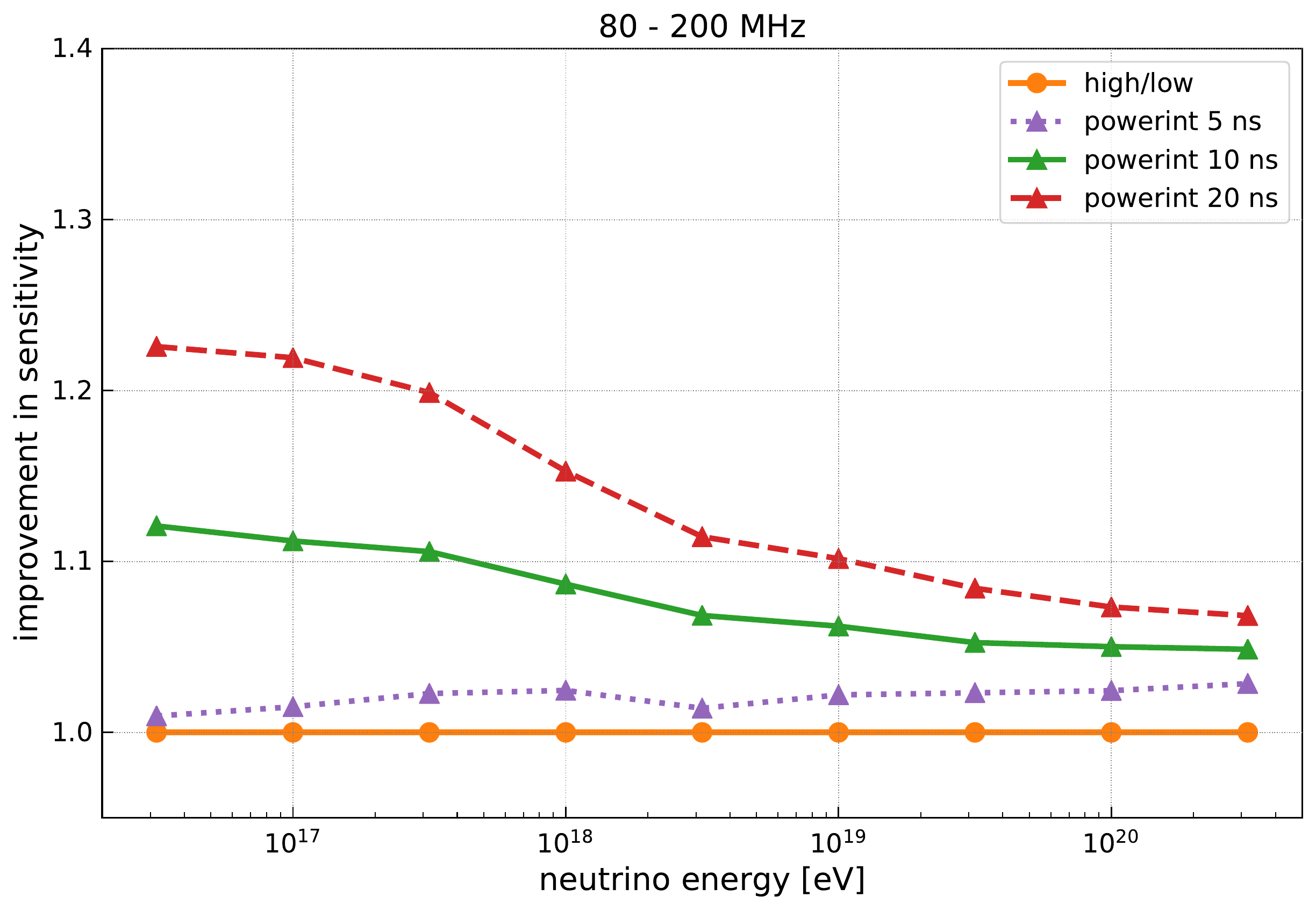}
    \caption{Improvement in sensitivity in the \SI{80}{MHz} to \SI{200}{MHz} band using the power integration trigger with respect to the high/low trigger scheme.}
    \label{fig:sensitivity_LPDA_200MHz}
\end{figure}

In Fig.~\ref{fig:sensitivity_LPDA_200MHz}, the different trigger schemes are compared for the best performing bandwidth of \SI{80}{MHz} to \SI{200}{MHz}. We find that the power integration trigger with the largest integration window of \SI{20}{ns} performs best and gives an additional boost in expected neutrino sensitivity of 15\% at \SI{e18}{eV}. The power integration trigger with a \SI{5}{ns} window performs very similar to the high/low trigger. 

\subsection{Discussion}
In all studied cases, the reduction in trigger bandwidth leads to a substantial increase in neutrino sensitivity. As visible from Fig.~\ref{fig:freq_spetrum}, the exact choice of trigger bandwidth can likely be further finetuned to optimally match different antenna responses. Other practical considerations such as hardware constraints on the implementation of the filters (e.g. the order of the filter) or site specific noise conditions that might require filtering of certain frequencies also need to be taken into account. For the high/low trigger, we explicitly tested the impact of a lower order lowpass filter which would be easier to implement on a chip. We found that a 1 pole filter works as well as the 10th order Butterworth lowpass filter that was used so far. Naively we would assume that the performance gets worse with a lower order because the filter lets through more high frequencies where the signal-to-noise ratio is worse. However, a feature of the high/low trigger works against that. In Fig.~\ref{fig:trigger_rate} left, we observed that the threshold increases with decreasing bandwidth because a downward fluctuation following an upward fluctuation becomes more likely. Lowering the order of the lowpass filter reduces this effect and results in a smaller threshold for the same bandwidth. These two effect cancel each other and make the gain in neutrino sensitivity of a 1 pole \SI{200}{MHz} lowpass filter comparable to a 10 pole \SI{200}{MHz} lowpass filter. 

We also tested if the neutrino sensitivity can be further improved by further restricting the bandwidth but found only a small additional effect. For the LPDA antennas of the ARIANNA detector we found that a bandwidth of \SI{80}{MHz} to \SI{150}{MHz} yields a slight additional improvement in neutrino sensitivity of a few percent which is what is being used in the proposed ARIANNA-200 detector \cite{ARIANNA200}. However, reducing the bandwidth further to \SI{80}{MHz} to \SI{100}{MHz} reduces the neutrino sensitivity. 

The sensitivity can be further improved by optimizing the trigger scheme. A power integration trigger with an optimal integration width performs better than a high/low trigger. The optimal integration window depends on the dispersion of the signal. In case of the relatively large dispersion of LPDA antennas, an integration window of \SI{20}{ns} performs best. For a different antenna type and/or dispersion of the rest of the signal chain, the optimal integration window might change. Thus, the general recommendation for a future Askaryan detector is to implement a power integration trigger where the integration window is optimized to the dispersion of the detector. 

A potential caveat of this study is the simulation of noise. If a significant non-thermal noise contribution is present, a certain trigger scheme might be more sensitive to this type of noise. For example, a low amplitude but long noise pulse will be most efficiently detected with a long integration window. Similarly, sharp fluctuations might be most efficiently detected by a high/low trigger. 
Therefore, it is recommended to confirm these findings using experimentally obtained noise data, e.g., by periodic readouts of the detector. 

As already discussed in the previous section, these findings also apply for an interferometric phased array trigger.
Thus, also for this trigger scheme, the sensitivity to neutrinos can be further optimized by restricting the bandwidth where details will depend on the dispersion of the antenna and signal chain. These optimizations were already performed for the interferometric phased array trigger system of the Radio Neutrino Observatory in Greenland (RNO-G) \cite{RNOGwhitepaper} where the first detector stations are expected to be installed in summer 2021. 
A smaller bandwidth also comes with the additional practical benefit of smaller digitization rates. The phased array requires a real time digitization. Thus, reducing digitization speeds from several GHz to \SI{400}{MHz} (which corresponds to the Nquist frequency at \SI{200}{MHz}) will save cost, power and data rates. 

\section{Conclusions}
Ultra-high energy (UHE) neutrinos can be measured cost-efficiently with an in-ice array of radio antennas that detect the Askaryan radiation following a neutrino interaction in the ice. The frequency content of these Askaryan signals ranges from around hundred MHz to a GHz. The technology has been successfully explored in test-bed arrays, and large detector arrays are currently in the planning and construction phase. The low flux of UHE neutrinos makes it vital to optimize the sensitivity of each detector station as much as possible which we do in this paper by systematically studying different trigger schemes and comparing approaches of current test-bed arrays. 

A method was presented to compare different trigger schemes in terms of their sensitivity to Askaryan signals.
We used this method to systematically compare the currently used trigger schemes, a high/low amplitude threshold trigger and a power integration trigger with various integration windows. We find that the optimal choice of trigger depends on the shape of the signal which is primarily determined by the detector response.
The bandwidth and dispersion of the detector were identified as relevant detector quantities.  We find that a power integration trigger where the integration window is optimized to the duration of the signal pulse typically performs best. For a small trigger bandwidth of \SI{80}{MHz} to \SI{200}{MHz}, the performance of the different trigger schemes become similar with a slight preference for a power integration trigger with optimized integration window. 

The impact of the trigger bandwidth on the sensitivity of an Askaryan detector to neutrinos was studied. We find that reducing the bandwidth from \SI{80}{MHz} to \SI{800}{MHz} that is currently being used to a much smaller bandwidth of \SI{80}{MHz} to \SI{200}{MHz} increases the sensitivity substantially. For the high/low trigger we find an increase of 50\% between the relevant neutrino energies of \SI{e17}{eV} and \SI{e18}{eV}. 

The findings also apply to the interferometric phased array trigger concept which produces a synthesised waveform with increased signal-to-noise ratio by coherently adding the signals from multiple antennas. The same trigger optimizations discussed here can be applied to the synthesised waveform.

These results give a clear recommendation for the trigger scheme of future detectors. The best sensitivity to neutrinos can be achieved with a restriction of the trigger bandwidth to low frequencies. 
We note that it is still advisable to record the full waveform with a larger bandwidth for event reconstruction purposes. The recommendation only applies to the trigger bandwidth. 
The sensitivity can be further improved by using a power-integration trigger where the integration window is matched to the dispersion of the detector. However, as the additional gain is modest compared to the gain from restricting the bandwidth, other considerations are likely of equal relevance, such as the effort of implementing and calibrating the trigger scheme in hardware or potential site-specific non-thermal noise which a certain trigger scheme might be more susceptible to than the other. 

In this work, only the first trigger stage was considered. To reduce the burden on input/output speeds, communication bandwidth, data storage and computing requirements, it is advisable to further reduce the trigger rate with secondary trigger stages that reject thermal noise while keeping Askaryan signals with high efficiency. Several approaches are currently being studied, among others neural networks that have the advantage of fast and constant processing speed and would allow to filter data in real-time. 

\section*{Acknowledgements}
The authors would like to thank their colleagues from the radio neutrino community for the lively discussions and input on this manuscript. We thank Allan Hallgren for providing detailed feedback on the manuscript. The author acknowledge funding from the German Research Foundation (DFG) under grant GL 914/1-1.

\begin{appendix}

\section{Treatment of thermal noise in NuRadioMC simulation}
\label{sec:noise}
Adding thermal noise to the Askaryan signals increased the trigger probability on average, and it is thus important to include, but comes with the technical challenge to not count thermal noise fluctuation as detected neutrinos. 
This a technical detail of the simulation but still worth reporting as it has been common practice to use a noiseless waveform
to calculate if a simulated neutrino interaction triggered the detector. For large enough trigger thresholds, adding thermal noise to the Askaryan signal has little influence when calculating the trigger condition. However, at the low trigger thresholds considered here (that correspond to a trigger rate of \SI{100}{Hz}) we find that adding noise has significant influence and an overall positive effect on the estimated neutrino sensitivity. 

Running a simulation with noise comes with the complication to make sure that pure thermal noise fluctuations are never counted as a neutrino observation. As already calculated before, at a thermal noise trigger rate of \SI{100}{Hz} and a trace length of \SI{256}{ns}, every 40 thousandth noise realization will fulfill the trigger condition. Especially at low neutrino energies, we often need to simulate more than 10,000 neutrino interactions for one triggered event. Thus, the likelihood for triggering on a noise fluctuation quickly becomes as likely as triggering on the Askaryan signal. In NuRadioMC we solve this problem by reducing the number of events for which the trigger is being simulated. This is achieved by several cuts that reject events if we already know that it will not produce a large enough signal, e.g., by rejecting events where the electric field amplitude at the antennas is already too low. As an additional cross check, a second simulation is performed where the Askaryan signal is set to zero before adding noise. Thus, the trigger is evaluated purely on thermal noise and we make sure that the number of triggered events is negligible compared to the standard simulation. 

The increase of sensitivity when simulating with thermal noise is strongest for a large bandwidth and leads to a 30\% to 40\% increase in predicted sensitivity at \SI{e18}{eV} depending on the trigger scheme. At a small bandwidth of \SI{80}{MHz} - \SI{200}{MHz}, the effect is much smaller with around 5\% increase in predicted sensitivity at \SI{e18}{eV}.
This is because at high bandwidth and in the presence of dispersion, the signal waveform exhibits multiple oscillations. Thus, thermal noise has several chances of shifting the waveform above threshold (cf. Fig.~\ref{fig:pulse_example} left vs. right).
At lower energies the effect gets stronger because the signals are more often close to the threshold. Correspondingly, the effect decreases towards higher energies.

\end{appendix}

\bibliographystyle{JHEP}
\bibliography{bib}

\providecommand{\href}[2]{#2}\begingroup\raggedright\begin{thebibliography}{10}

\bibitem{PhysRevLett.117.241101}
{\scshape IceCube} collaboration, \emph{Constraints on ultrahigh-energy
  cosmic-ray sources from a search for neutrinos above 10 {PeV} with
  {IceCube}},
  \href{https://doi.org/10.1103/PhysRevLett.117.241101}{\emph{Physical Review
  Letters} {\bfseries 117} (2016) 241101} Erratum 119 (2017) 259902.

\bibitem{DecadalWhitePaper}
M.~Ackermann et~al., \emph{{Astrophysics Uniquely Enabled by Observations of
  High-Energy Cosmic Neutrinos}}, {\emph{Bulletin of the American Astronomical
  Society} {\bfseries 51} (2019) 185}
  [\href{https://arxiv.org/abs/1903.04334}{{\ttfamily 1903.04334}}].

\bibitem{DecadalWhitePaper2}
M.~Ackermann et~al., \emph{{Fundamental Physics with High-Energy Cosmic
  Neutrinos}}, {\emph{Bulletin of the American Astronomical Society} {\bfseries
  51} (2019) 215} [\href{https://arxiv.org/abs/1903.04333}{{\ttfamily
  1903.04333}}].

\bibitem{Aartsen:2013rt}
{\scshape IceCube} collaboration, \emph{{Measurement of South Pole ice
  transparency with the IceCube LED calibration system}},
  \href{https://doi.org/10.1016/j.nima.2013.01.054}{\emph{Nucl. Instrum. Meth.}
  {\bfseries A711} (2013) 73}
  [\href{https://arxiv.org/abs/1301.5361}{{\ttfamily 1301.5361}}].

\bibitem{Askaryan}
G.~A. {Askar'yan}, \emph{{Coherent Radio Emission from Cosmic Showers in Air
  and in Dense Media}}, {\emph{Soviet Journal of Experimental and Theoretical
  Physics} {\bfseries 21} (1965) 658}.

\bibitem{ARIANNA2015}
{\scshape ARIANNA} collaboration, \emph{{Design and Performance of the ARIANNA
  HRA-3 Neutrino Detector Systems}},
  \href{https://doi.org/10.1109/TNS.2015.2468182}{\emph{IEEE Trans. Nucl. Sci.}
  {\bfseries 62} (2015) 2202}
  [\href{https://arxiv.org/abs/1410.7369}{{\ttfamily 1410.7369}}].

\bibitem{ARIA}
{\scshape ARIANNA} collaboration, \emph{{Targeting ultra-high energy neutrinos
  with the ARIANNA experiment}},
  \href{https://doi.org/https://doi.org/10.1016/j.asr.2019.06.016}{\emph{Advances
  in Space Research} {\bfseries 64} (2019) 2595}.

\bibitem{ARA}
{\scshape ARA} collaboration, \emph{{Performance of two Askaryan Radio Array
  stations and first results in the search for ultrahigh energy neutrinos}},
  \href{https://doi.org/10.1103/PhysRevD.93.082003}{\emph{Physical Review D}
  {\bfseries 93} (2016) 082003}
  [\href{https://arxiv.org/abs/1507.08991}{{\ttfamily 1507.08991}}].

\bibitem{Anker_2020}
{\scshape ARIANNA} collaboration, \emph{A search for cosmogenic neutrinos with
  the {ARIANNA} test bed using 4.5 years of data},
  \href{https://doi.org/10.1088/1475-7516/2020/03/053}{\emph{Journal of
  Cosmology and Astroparticle Physics} {\bfseries 03} (2020) 053}
  [\href{https://arxiv.org/abs/1909.00840}{{\ttfamily 1909.00840}}].

\bibitem{ARA2019}
{\scshape ARA} collaboration, \emph{Constraints on the diffuse flux of
  ultra-high energy neutrinos from four years of {Askaryan Radio Array} data in
  two stations}, \href{https://doi.org/10.1103/PhysRevD.102.043021}{\emph{Phys.
  Rev. D} {\bfseries 102} (2020) 043021}
  [\href{https://arxiv.org/abs/1912.00987}{{\ttfamily 1912.00987}}].

\bibitem{ARAprogress}
{\scshape ARA} collaboration, \emph{Design and performance of an
  interferometric trigger array for radio detection of high-energy neutrinos},
  \href{https://doi.org/10.1016/j.nima.2019.01.067}{\emph{Nuclear Instruments
  and Methods in Physics Research A} {\bfseries 930} (2019) 112}
  [\href{https://arxiv.org/abs/1809.04573}{{\ttfamily 1809.04573}}].

\bibitem{Barwick2014}
{\scshape ARIANNA} collaboration, \emph{Design and performance of the {ARIANNA}
  {HRA}-3 neutrino detector systems},
  \href{https://doi.org/10.1109/TNS.2015.2468182}{\emph{{IEEE} Transactions on
  Nuclear Science} {\bfseries 62} (2015) 2202}
  [\href{https://arxiv.org/abs/1410.7369}{{\ttfamily 1410.7369}}].

\bibitem{RNOGwhitepaper}
{\scshape RNO-G} collaboration, \emph{Design and sensitivity of the radio
  neutrino observatory in {Greenland} {(RNO-G)}}, {\emph{Journal of
  Instrumentation} {\bfseries 16} (2021) P03025}
  [\href{https://arxiv.org/abs/2010.12279}{{\ttfamily 2010.12279}}].

\bibitem{GlaserICRC2019}
{C. Glaser for the ARIANNA collaboration}, \emph{{Neutrino direction and energy
  resolution of Askaryan detectors}},  in \emph{36th International Cosmic Ray
  Conference (ICRC2019)}, vol.~36 of \emph{International Cosmic Ray
  Conference}, p.~899, July, 2019,
  \href{https://arxiv.org/abs/1911.02093}{{\ttfamily 1911.02093}}.

\bibitem{ARIANNA2020Polarization}
{\scshape ARIANNA} collaboration, \emph{Probing the angular and polarization
  reconstruction of the {ARIANNA} detector at the {South Pole}},
  \href{https://doi.org/10.1088/1748-0221/15/09/P09039}{\emph{Journal of
  Instrumentation} {\bfseries 15} (2020) P09039}
  [\href{https://arxiv.org/abs/2006.03027}{{\ttfamily 2006.03027}}].

\bibitem{Gen2WhitePaper}
{\scshape IceCube-Gen2} collaboration, \emph{{IceCube-Gen2: The Window to the
  Extreme Universe}},  \href{https://arxiv.org/abs/2008.04323}{{\ttfamily
  2008.04323}}.

\bibitem{Barwick2017}
{\scshape ARIANNA} collaboration, \emph{{Radio detection of air showers with
  the ARIANNA experiment on the Ross Ice Shelf}},
  \href{https://doi.org/10.1016/j.astropartphys.2017.02.003}{\emph{Astropart.
  Phys.} {\bfseries 90} (2017) 50}.

\bibitem{GarciaFernandez2020}
D.~García-Fernández, C.~Glaser and A.~Nelles, \emph{The signatures of
  secondary leptons in radio-neutrino detectors in ice},
  \href{https://doi.org/10.1103/PhysRevD.102.083011}{\emph{Physical Review D}
  {\bfseries 102} (2020) 083011}
  [\href{https://arxiv.org/abs/2003.13442}{{\ttfamily 2003.13442}}].

\bibitem{ARIANNA200}
{\scshape ARIANNA} collaboration, \emph{{White Paper: ARIANNA-200 high energy
  neutrino telescope}},  \href{https://arxiv.org/abs/2004.09841}{{\ttfamily
  2004.09841}}.

\bibitem{WindTurbineICRC2019}
{A. Nelles for the ARIANNA Collaboration}, \emph{{A wind-turbine for autonomous
  stations for radio detection of neutrinos}}, {\emph{Proc. 36th ICRC 2019,
  Madison, Wisconsin, USA, PoS(ICRC2019)968} }.

\bibitem{NuRadioReco}
C.~Glaser, A.~Nelles, I.~Plaisier, C.~Welling, S.~W. Barwick,
  D.~García-Fernández et~al., \emph{{NuRadioReco}: A reconstruction framework
  for radio neutrino detectors},
  \href{https://doi.org/10.1140/epjc/s10052-019-6971-5}{\emph{The European
  Physical Journal C} {\bfseries 79} (2019) }
  [\href{https://arxiv.org/abs/1903.07023}{{\ttfamily 1903.07023}}].

\bibitem{mexicanhat}
``Wikipedia: {Mexican hat wavelet}.''
  \url{https://en.wikipedia.org/wiki/Mexican_hat_wavelet}.

\bibitem{ARIANNATimeDomain}
{\scshape ARIANNA} collaboration, \emph{{Time-domain response of the ARIANNA
  detector}},
  \href{https://doi.org/10.1016/j.astropartphys.2014.09.002}{\emph{Astroparticle
  Physics} {\bfseries 62} (2015) 139–151}.

\bibitem{AlvarezMuiz2000}
J.~Alvarez-Mu{\~{n}}iz, R.~A. V{\'{a}}zquez and E.~Zas, \emph{Calculation
  methods for radio pulses from high energy showers},
  \href{https://doi.org/10.1103/physrevd.62.063001}{\emph{Physical Review D}
  {\bfseries 62} (2000) }.

\bibitem{NuRadioMC}
C.~Glaser et~al., \emph{{NuRadioMC: Simulating the radio emission of neutrinos
  from interaction to detector}},
  \href{https://doi.org/10.1140/epjc/s10052-020-7612-8}{\emph{European Physical
  Journal} {\bfseries C80} (2020) 77}
  [\href{https://arxiv.org/abs/1906.01670}{{\ttfamily 1906.01670}}].

\bibitem{ZHS}
E.~Zas, F.~Halzen and T.~Stanev, \emph{Electromagnetic pulses from high-energy
  showers: Implications for neutrino detection},
  \href{https://doi.org/10.1103/PhysRevD.45.362}{\emph{Phys. Rev. D} {\bfseries
  45} (1992) 362}.

\bibitem{DnR2019}
{\scshape ARIANNA} collaboration, \emph{Neutrino vertex reconstruction with
  in-ice radio detectors using surface reflections and implications for the
  neutrino energy resolution},
  \href{https://doi.org/10.1088/1475-7516/2019/11/030}{\emph{Journal of
  Cosmology and Astroparticle Physics} {\bfseries 11} (2019) 030}
  [\href{https://arxiv.org/abs/http://arxiv.org/abs/1909.02677v1}{{\ttfamily
  http://arxiv.org/abs/1909.02677v1}}].

\end{thebibliography}\endgroup
\end{document}